\documentclass[twocolumn]{aastex631}

\begin{document}

\title{Discovery of the elusive carbonic acid (HOCOOH) in space}


\author[0000-0001-9629-0257]{Miguel Sanz-Novo}
\affiliation{Centro de Astrobiolog{\'i}a (CAB), INTA-CSIC, Carretera de Ajalvir km 4, Torrej{\'o}n de Ardoz, 28850 Madrid, Spain}
\affiliation{Computational Chemistry Group, Departamento de Química Física y Química Inorgánica, Universidad de Valladolid, E-47011 Valladolid, Spain}

\author[0000-0002-2887-5859]{V\'ictor M. Rivilla}
\affiliation{Centro de Astrobiolog{\'i}a (CAB), INTA-CSIC, Carretera de Ajalvir km 4, Torrej{\'o}n de Ardoz, 28850 Madrid, Spain}

\author[0000-0003-4493-8714]{Izaskun Jim\'enez-Serra}
\affiliation{Centro de Astrobiolog{\'i}a (CAB), INTA-CSIC, Carretera de Ajalvir km 4, Torrej{\'o}n de Ardoz, 28850 Madrid, Spain}

\author[0000-0003-4561-3508]{Jes\'us Mart\'in-Pintado}
\affiliation{Centro de Astrobiolog{\'i}a (CAB), INTA-CSIC, Carretera de Ajalvir km 4, Torrej{\'o}n de Ardoz, 28850 Madrid, Spain}

\author[0000-0001-8064-6394]{Laura Colzi}
\affiliation{Centro de Astrobiolog{\'i}a (CAB), INTA-CSIC, Carretera de Ajalvir km 4, Torrej{\'o}n de Ardoz, 28850 Madrid, Spain}

\author[0000-0003-3721-374X]{Shaoshan Zeng}
\affiliation{Star and Planet Formation Laboratory, Cluster for Pioneering Research, RIKEN, 2-1 Hirosawa, Wako, Saitama, 351-0198, Japan}

\author[0000-0002-6389-7172]{Andr\'es Meg\'ias}
\affiliation{Centro de Astrobiolog{\'i}a (CAB), INTA-CSIC, Carretera de Ajalvir km 4, Torrej{\'o}n de Ardoz, 28850 Madrid, Spain}

\author[0000-0001-6049-9366]{\'Alvaro L\'opez-Gallifa}
\affiliation{Centro de Astrobiolog{\'i}a (CAB), INTA-CSIC, Carretera de Ajalvir km 4, Torrej{\'o}n de Ardoz, 28850 Madrid, Spain}

\author[0000-0001-5191-2075]{Antonio Mart\'inez-Henares}
\affiliation{Centro de Astrobiolog{\'i}a (CAB), INTA-CSIC, Carretera de Ajalvir km 4, Torrej{\'o}n de Ardoz, 28850 Madrid, Spain}

\author[0000-0002-7387-9787]{Sarah Massalkhi}
\affiliation{Centro de Astrobiolog{\'i}a (CAB), INTA-CSIC, Carretera de Ajalvir km 4, Torrej{\'o}n de Ardoz, 28850 Madrid, Spain}

\author[0000-0002-4782-5259]{Bel\'en Tercero}
\affiliation{Observatorio Astron\'omico Nacional (OAN-IGN), Calle Alfonso XII, 3, 28014 Madrid, Spain}

\author[0000-0002-5902-5005]{Pablo de Vicente}
\affiliation{Observatorio de Yebes (OY-IGN), Cerro de la Palera SN, Yebes, Guadalajara, Spain}






\author[0000-0001-9281-2919]{Sergio Mart\'in}
\affiliation{European Southern Observatory, Alonso de C\'ordova 3107, Vitacura 763 0355, Santiago, Chile}
\affiliation{Joint ALMA Observatory, Alonso de C\'ordova 3107, Vitacura 763 0355, Santiago, Chile}

\author[0000-0001-7535-4397]{David San Andr\'es}
\affiliation{Centro de Astrobiolog{\'i}a (CAB), INTA-CSIC, Carretera de Ajalvir km 4, Torrej{\'o}n de Ardoz, 28850 Madrid, Spain}

\author[0009-0009-5346-7329]{Miguel A. Requena-Torres}
\affiliation{University of Maryland, College Park, ND 20742-2421 (USA)}
\affiliation{Department of Physics, Astronomy and Geosciences, Towson University, Towson, MD 21252, USA}

\begin{abstract}

After a quarter century since the detection of the last interstellar carboxylic acid, acetic acid (CH$_3$COOH), we report the discovery of a new one, the \textit{cis-trans} form of carbonic acid (HOCOOH), toward the Galactic Center molecular cloud G+0.693-0.027. HOCOOH stands as the first interstellar molecule containing three oxygen atoms and also the third carboxylic acid detected so far in the interstellar medium. Albeit the limited available laboratory measurements (up to 65 GHz), we have also identified several pairs of unblended lines directly in the astronomical data (between 75-120 GHz), which allowed us to slightly improve the set of spectroscopic constants. We derive a column density for \textit{cis-trans} HOCOOH of $N$ = (6.4 $\pm$ 0.4) $\times$ 10$^{12}$ cm$^{-2}$, which yields an abundance with respect to molecular H$_2$ of 4.7 $\times$ 10$^{-11}$. Meanwhile, the extremely low dipole moment (about fifteen times lower) of the lower-energy conformer, \textit{cis-cis} HOCOOH, precludes its detection. We obtain an upper limit to its abundance with respect to H$_2$ of $\leq$ 1.2 $\times$10$^{-9}$, which suggests that \textit{cis-cis} HOCOOH might be fairly abundant in interstellar space, although it is nearly undetectable by radio astronomical observations. We derive a \textit{cis-cis}/\textit{cis-trans} ratio $\leq$ 25, consistent with the smaller energy difference between both conformers compared with the relative stability of \textit{trans-} and \textit{cis}-formic acid (HCOOH). Finally, we compare the abundance of these acids in different astronomical environments, further suggesting a relationship between the chemical content found in the interstellar medium and the chemical composition of the minor bodies of the Solar System, which could be inherited during the star formation process.




\end{abstract}

\section{Introduction} 
\label{sec:intro}


In recent times, significant endeavors have been made to study the molecular complexity of the interstellar medium (ISM). Astronomical observations have proven that interstellar chemistry can generate diverse complex organic molecules (or COMs, defined as carbon-based molecules comprised of 6 or more atoms; \citealt{Herbst2020}), highlighting several building blocks of key biomolecules (see \citealt{McGuire22census} for a recent molecular census).



Among all COMs, carboxylic acids occur widely in nature and are considered precursors of many relevant prebiotic molecules, such as amino acids and lipids \citep{Ehrenfreund2001,Georgiou2014,Zhu2018}. Almost fifty years ago, \cite{Zuckerman:1971de} reported the detection of the first interstellar carboxylic acid, formic acid (HCOOH), in the star-forming region Sgr B2. It took more than twenty years to detect acetic acid (CH$_3$COOH), which was also primarily identified toward Sgr B2 \citep{Mehringer:1997vk}. Currently, both acids represented, so far, the only carboxylic acids conclusively detected in the ISM (see e.g., \citealt{irvine1990, Liu2002,Remijan:2003wf,requena-torres_organic_2006,Cuadrado:2016hp,lefloch_l1157-b1_2017,rivilla_chemical_2017,jorgensen2018,Tercero2018,rodriguez-almeida2021a}).


A more exotic yet well-known carboxylic acid is carbonic acid (H$_2$CO$_3$ or HOCOOH), an hydroxy-derivative of formic acid, which appears as an auspicious astronomical candidate. This molecule plays a relevant role in various biological and geochemical processes \citep{Adamczyk2009,Loerting2000}, highlighting its implications in the global carbon cycle \citep{Jones2014,Wang2016}, and in particular in the anthropogenic carbon and ocean pH \citep{Caldeira2003,Ioppolo2021}. Moreover, the presence of this molecule has also been suggested in different astronomical environments such as the Galilean icy moons \citep{Delitsky1998,Jones2014,Bennett2014}, on Mercury’s north polar region \citep{Delitsky2017}, or even on the surface and/or atmosphere of Mars \citep{Strazzulla1996}. In fact, it has been proposed to be formed on the icy mantles of dust grains, which contain vast amounts of H$_2$O and CO$_2$ \citep{Moore2001,Hage1998,Zheng2007,Oba2010,Ioppolo2021}. Nevertheless, HOCOOH still awaited detection in the ISM.

Additionally, carboxylic acids are deeply ingrained in the chemical reservoir of carbonaceous chondrite meteorites \citep{Cooper1992,Sephton2002,Pizzarello2010,Glavin2010,pizzarello2012}, and comets such as the 67P/Churyumov–Gerasimenko \citep{Altwegg:2016ck}, which further motivates the astrophysical community to search for related species in the ISM. Very recently, diverse COMs including HCOOH, CH$_3$COOH and a variety of amino acids, have been detected through mass spectrometry in the samples of the near-Earth carbonaceous (C-type) asteroid 162173 Ryugu, gathered and transported to Earth by the Hayabusa2 spacecraft \citep{Naraoka2023}.

The presence of prebiotic COMs within extraterrestrial material thus firmly suggests the existence of carboxylic acids of increasing complexity in the ISM, including amino acid-related species. Within this context, considerable efforts have been devoted to hunt for other acids such as propenoic or acrylic acid (CH$_2$CHCOOH; \citealt{Alonso2015A}), propanoic acid (CH$_3$CH$_2$COOH; \citealt{Ilyushin2021}), cyanoacetic acid (CH$_2$CNCOOH; \citealt{Sanz-Novo21}), glycolic acid (CH$_2$OHCOOH; \citealt{Blom1981,Jimenez-Serra20}), hydantoic acid (C(O)OHCH$_2$NHC(O)NH$_2$; \citealt{Kolesnikova19}), and glycine (CH$_2$(NH$_2$)C(O)OH; \citealt[][]{combes1996,Hollis03,Kuan:2003yt,Cunningham07,Jones07,Jimenez-Serra16,Jimenez-Serra20}), whose identification in the ISM remains elusive \citep{Snyder05}.

In this work, we present the discovery in the ISM of carbonic acid, which has been detected toward the chemically-rich G+0.693-0.027 molecular cloud, located in the Galactic Center. We performed the search for the two most stable conformers of the molecule, previously characterized in the laboratory by \cite{Mori2009,Mori2011}, and we report the conclusive detection of the higher-in-energy \textit{cis-trans} form and derive an upper limit to the column density for the \textit{cis-cis} conformer. Additionally, we discuss about their relative ratio, including a comparison with the well-known interstellar carboxylic acids: formic acid and acetic acid toward several astronomical environments.

\section{Observations} \label{sec:obs}

We searched for the \textit{cis-cis} and \textit{cis-trans} conformers of HOCOOH toward the Galactic Center molecular cloud G+0.693-0.027 (hereafter G+0.693). This prominent astronomical source is one of the main repositories of COMs in the Milky Way. To date, the analysis of the unbiased spectral survey conducted toward G+0.693 has resulted in the first interstellar detections of more than a dozen COMs (see, e.g., \citealt{rivilla2019b,rivilla2020b,rivilla2021a,rivilla2021b,rivilla2022a,rivilla2022b,rodriguez-almeida2021a,rodriguez-almeida2021b,jimenez-serra2022,zeng2021,zeng2023}).

This spectral survey was performed with the Yebes 40m (Guadalajara, Spain) and the IRAM 30m (Granada, Spain) telescopes, and its sensitivity has recently been enhanced compared to previous works (e.g., \citealt{rivilla2021a,rodriguez-almeida2021a,jimenez-serra2022}). For the observations, we employed the position switching mode, centered at $\alpha$ = $\,$17$^{\rm h}$47$^{\rm m}$22$^{\rm s}$, $\delta$ = $\,-$28$^{\circ}$21$^{\prime}$27$^{\prime\prime}$, with the off position shifted by $\Delta\alpha$~=~$-885$$^{\prime\prime}$ and $\Delta\delta$~=~$290$$^{\prime\prime}$. 

The new Yebes 40m observations (project 21A014; PI Rivilla) comprised different observing runs performed between March 2021 and March 2022. We employed the ultra broadband Nanocosmos Q-band (7\,mm) HEMT receiver that allows broadband observations (18.5\,GHz) in two linear polarizations (\citealt{tercero2021}). The backends are 16 Fast Fourier Transform Spectrometers (FFTS) which provide a raw channel width of 38\,kHz. Two distinct spectral setups were used, centered at 41.4 and 42.3\,GHz, respectively and covering the whole $Q$-band (frequency range: 31.07$-$50.42\,GHz). We initially examined and reduced the data employing a Python-based script \citep{Megias2023}\footnote{\url{https://github.com/andresmegias/gildas-class-python/}} that uses the \textsc{Class} module within the \textsc{Gildas} package. Hence, a systematic approach was followed for each observing day. We automatically fitted baselines using an iterative method that first masks the more visible lines using \textit{sigma-clips} and then applies rolling medians and averages, interpolating the masked regions with splines. Afterward, the spectra were combined, averaged, and subsequently imported to \textsc{Madcuba} \citep{martin2019}. Note that we compared the spectra obtained using the different frequency setups to inspect possible line contamination from spurious lines and other technical artefacts. We finally smoothed the resulting spectra to 256\,kHz, which translates into velocity resolutions of 1.5$-$2.5 km s$^{-1}$ in the range observed. We checked flux density consistency between the new data and the previous Yebes 40m survey (e.g. \citealt{zeng2020}) obtaining a good match (within a 5$\%$ level). We have achieved an extremely high level of sensitivity with rms noise levels lying between 0.25$-$0.9 mK across the whole $Q$-band at this spectral resolution in antenna temperature ($T$$\mathrm{_A^*}$) scale, as the molecular emission toward G+6.693 is extended over the beam \citep{requena-torres_organic_2006,requena-torres_largest_2008,zeng2018,zeng2020}. The half power beam width (HPBW) of the telescope varies between $\sim$35$-$55$^{\prime\prime}$ across the frequency range covered \citep{tercero2021}.

\begin{center}
\begin{figure*}[ht]
     \centerline{\resizebox{1.02
     \hsize}{!}{\includegraphics[angle=0]{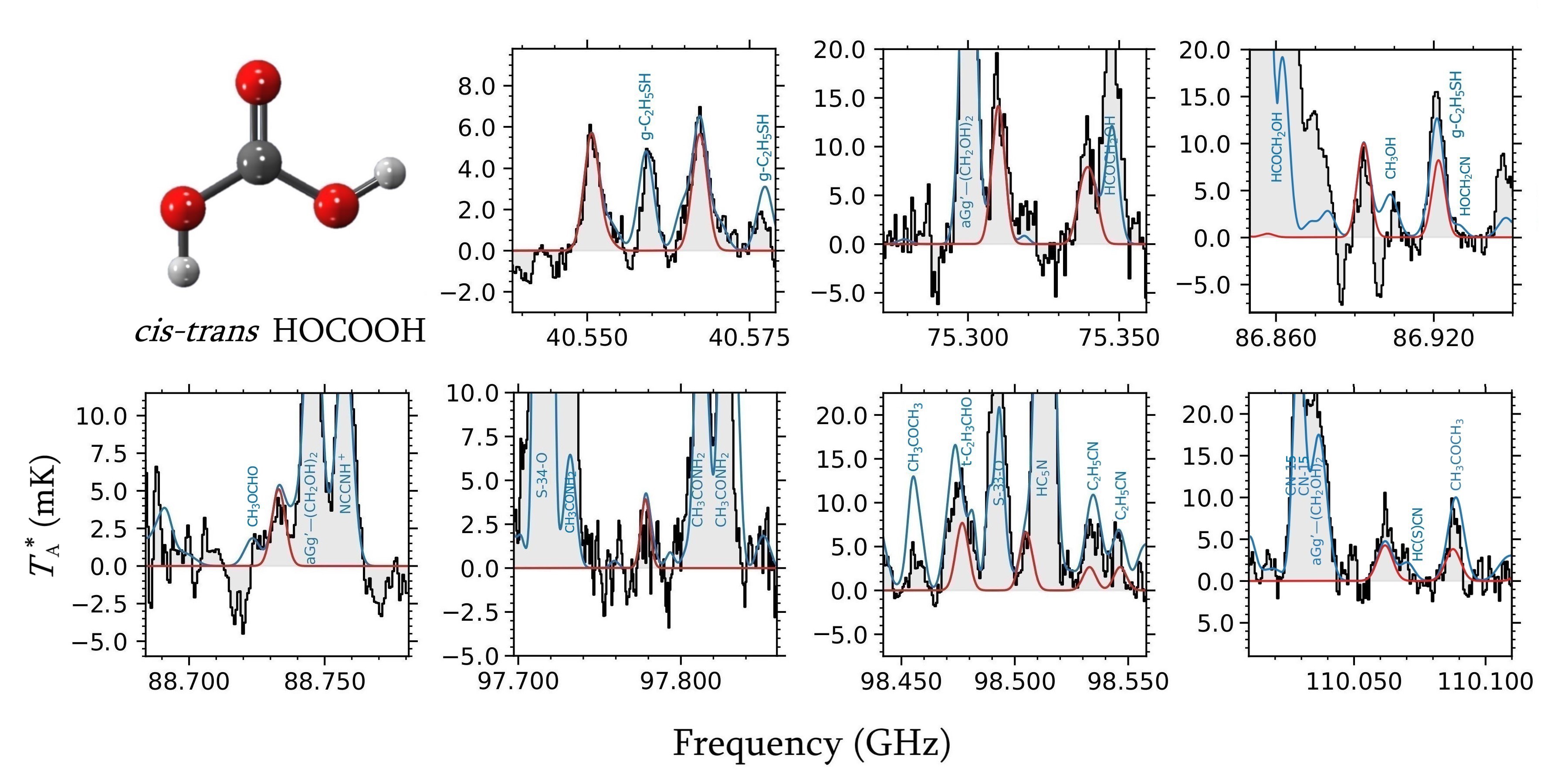}}}
     \caption{Selected transitions of \textit{cis-trans} HOCOOH identified toward the G+0.693–0.027 molecular cloud (listed in Table \ref{tab:h2co3unblended}). The result of the best LTE fit is shown with a red solid line, while the blue line shows the expected molecular emission from all the molecular species identified to date in our survey. The observed spectra are plotted as gray histograms. The structure of \textit{cis-trans} HOCOOH is also depicted (carbon atoms in grey; oxygen atoms are in red, and hydrogen atoms in white).}
\label{f:LTEspectrum}
\end{figure*}
\end{center}

Regarding the new IRAM 30m observations (project 123-22, PI Jim\'enez-Serra), they were performed between February 1$-$18 2023. We used several frequency setups to cover different spectral ranges of the E0 (3 mm) and E1 (2 mm) bands of the multi-band mm-wave receiver named Eight MIxer Receiver (EMIR). An initial spectral resolution of 195 kHz was achieved by using the Fast Fourier Transform Spectrometer (FTS200). Each setup was slightly shifted in frequency in order to spot possible contamination of spurious lines from the image band. We covered three frequency ranges: 83.2$-$115.41 GHz, 132.28$-$140.39 and 142$-$173.81 GHz, and the HPBW is $\sim$14$^{\prime\prime}$$-$29$^{\prime\prime}$. We imported the spectra from \textsc{Class} to \textsc{Madcuba} and subsequently compared the line intensities of the new astronomical data with those of previous observations (e.g. \citealt{rodriguez-almeida2021a}), obtaining an excellent consistency (within a 5$\%$ level). Afterward, we averaged the above data within \textsc{Madcuba} by weighing the spectra according to their rms noise level. We then smoothed the resulting spectra to 615\,kHz, which corresponds to velocity resolutions of 1.0$-$2.2 km s$^{-1}$ in the observed frequency range. All in all, we reached noise levels between 0.5$-$2.5 mK at 3 mm, and 1.0$-$1.6 mK at 2 mm. For the frequency ranges that are not covered by the new observations, we kept our previous IRAM 30m survey (further details are reported elsewhere; e.g., \citealt{rodriguez-almeida2021a,rivilla2022b}).

\section{Analysis and results} 
\label{sec:res}
\subsection{Rotational spectroscopy of carbonic acid} 
\label{subsec:spec}
\begin{figure}
\centerline{\resizebox{1.05\hsize}{!}{\includegraphics[angle=0]{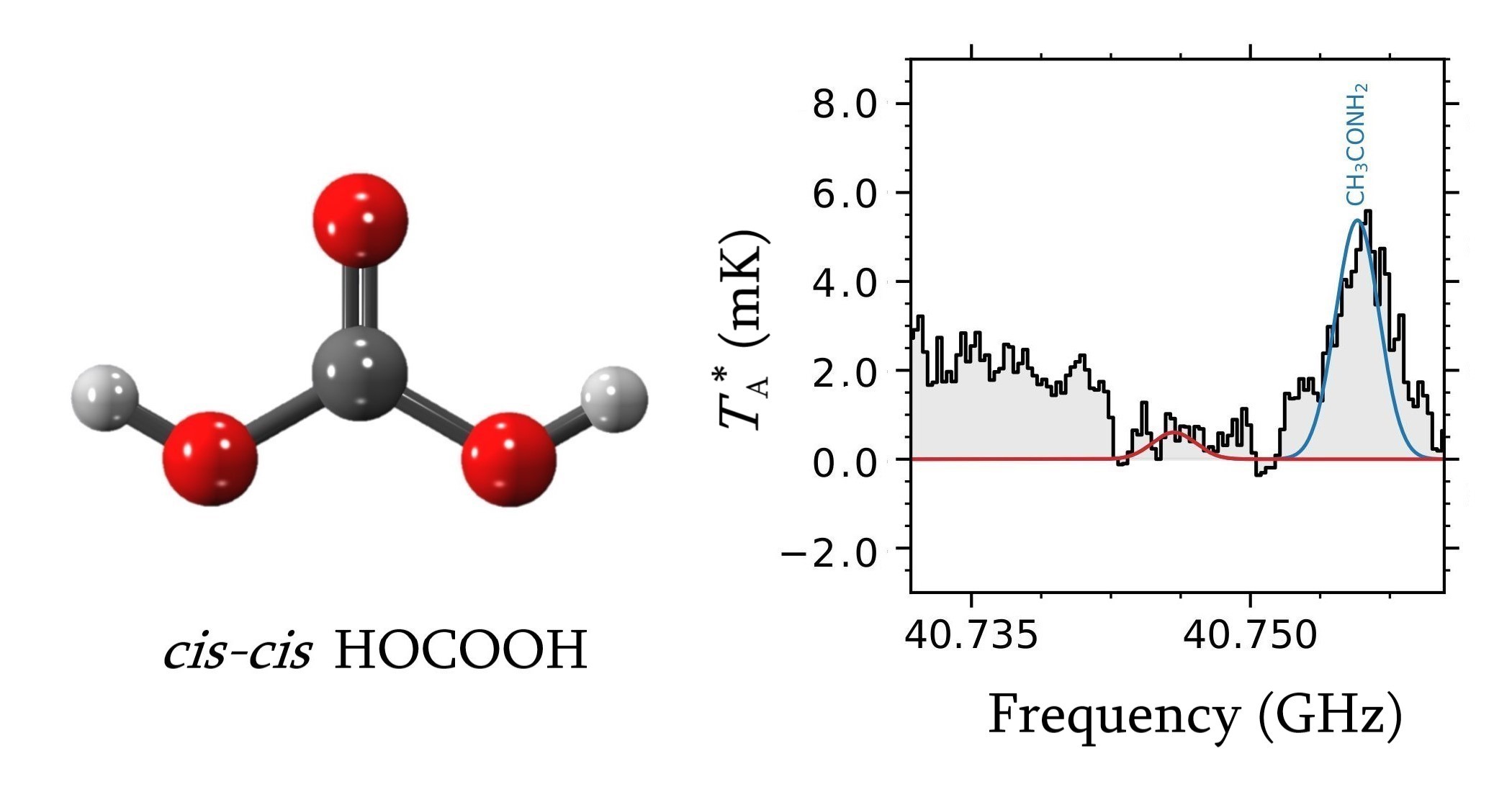}}}
\caption{LTE simulation of the \textit{cis-cis} HOCOOH emission at the 3$\sigma$ upper limit column density derived toward G+0.693 using the physical parameters shown in Table \ref{tab:comparisonacids} (in red) together with the expected molecular emission from all the molecular species identified to date in our survey (in blue), both overlaid on the observations (black line and in gray histogram). The features depicted correspond to the 3$_{1,3}-$2$_{0,2}$ rotational transition. The structure of \textit{cis-cis} HOCOOH is also shown.}
\label{f:LTEspectrumciscisbis}
\end{figure}

The spectroscopic data of HOCOOH were obtained from previous Fourier-transform microwave (FTMW) and double resonance measurements carried out by \citet{Mori09,Mori2011}. The conformational landscape of this carboxilic acid shows three distinct conformational isomers or conformers, \textit{cis-cis} (global minimum in energy), \textit{cis-trans}, and \textit{trans-trans} (located at 1.74 and 10.9 kcal mol$^{-1}$ higher in energy, respectively), although only the \textit{cis-cis} and the \textit{cis-trans} forms have been measured in the laboratory. These two HOCOOH conformers are asymmetric tops close to the oblate limit (see Figure \ref{f:LTEspectrum} and \ref{f:LTEspectrumciscisbis}). Thus, according to the dipole moment selection rules, \textit{a}- and \textit{b}-type lines are permitted for the \textit{cis-trans} conformer ($\mu_a$ = 1.0 D and $\mu_b$ = 2.9 D, computed at the CCSD(T)/cc-pVQZ level, \citealt{Mori09}), while only \textit{b}-type spectrum is observable for the \textit{cis-cis} form. However, the latter exhibit an extremely low dipole moment ($\mu_b$ = 0.2 D at the CCSD(T)/cc-pVQZ level, \citealt{Mori2011}), which hampered its detection in the first spectroscopic characterization of the isolated molecule \citep{Mori09}. Consequently, we can foresee that only the direct identification of the \textit{cis-trans} conformer will be feasible in the ISM, since its \textit{b}-type dipole moment component is almost fifteen times larger than that of \textit{cis-cis} HOCOOH.

So far, only seven \textit{b}-type \textit{R}- and \textit{Q}-branch transitions have been reported in the literature for \textit{cis-cis} HOCOOH in the 6.1-41 GHz frequency range \citep{Mori09}, while a total of 25 transitions, including different \textit{a}-type and \textit{b}-type lines, have been measured for the \textit{cis-trans} conformer from 4.8 to 65 GHz \citep{Mori2011}. We employed all the available transition frequencies to perform a least-square fit to a semirigid-rotor Hamiltonian for each conformer, separately (Watson’s A-reduced Hamiltonian in $I$$^{r}$-representation). The resulting spectroscopic constants are listed in Table \ref{t:mwdata}. We then extrapolated these results to higher frequencies (up to $\sim$120 GHz) in order to prepare separate line catalogs to search for both conformers in the ultra-sensitive unbiased spectral survey toward G+0.693.

After performing the search, several lines of \textit{cis-trans} HOCOOH were straightforwardly identified in the astronomical data (i.e, the detected transitions at $\sim$40 GHz were perfectly fitted by the predictions based exclusively on \citealt{Mori09}; see Section \ref{subsec:detection} for more detailed information). Nevertheless, it is worth noticing that systematic discrepancies up to $\sim$5 km s$^{-1}$ were initially found for the higher-in-frequency astronomical lines (e.g. those at $\sim$110 GHz) when compared to our predictions. As expected, this fact points to the need of higher order centrifugal distortion constants to perfectly reproduce the astronomical spectra, especially at frequencies higher than $\sim$90 GHz, since the available laboratory data is limited to 65 GHz. At this point, we employed pairs of $a$- and $b$-type transitions with $K$$_a$ = 0,1 (\textit{J} = 5, 6 and 8) and $K$$_a$ = 1,2 (\textit{J} = 7) that are merged into quadruply degenerated lines (four astronomical lines with excellent signal-to-noise ratio, S/N $>$ 10) to perform a global fit together with the previous laboratory measurements \citep{Mori09,Mori2011}. We treated the astronomical data using the Pickett's SPFIT/SPCAT program suite \citep{Pickett:1991cv} and fitted each observational line to a Gaussian profile to obtain its rest frequency. The weight of the purely astronomical lines was fixed to a much higher value than that used for the laboratory data; we employed 0.3 MHz that corresponds to $ \sim$1 km s$^{-1}$ at 100 GHz. The derived spectroscopic parameters for the global fit of \textit{cis-trans} HOCOOH are listed in the third column of Table~\ref{t:mwdata}. The sextic centrifugal distortion constant $\Phi_J$ was determined from the fit, which enabled us to improve the predictions for the higher \textit{J} transitions. Finally, we prepared a new catalog for the \textit{cis-trans} conformer (ctHOCOOH.cat; provided as Supporting Information in the common JPL SPFIT/SPCAT format and an example table, Table \ref{t:cat}, is also given in Appendix \ref{t:mwdata}.), which was used in the following subsection to confirm the detection and to derive the physical parameters of the molecule. This catalogue can be also directly employed to search for \textit{cis-trans} HOCOOH toward other astronomical sources.

\subsection{Detection of \textit{cis-trans} carbonic acid and search for the \textit{cis-cis} conformer} 
\label{subsec:detection}

Once we imported the spectroscopic catalog of HOCOOH into the \textsc{Madcuba} package \citep{martin2019}, the identification of the astronomical lines was carried out using the Spectral Line Identification and Modeling (SLIM) tool, which works under the assumption of Local Thermodynamic Equilibrium (LTE) excitation to generate the synthetic spectra.

\begin{table*}
\centering
\tabcolsep 3pt
\caption{Spectroscopic information of the selected unblended or slightly blended transitions of \textit{cis-trans} HOCOOH detected toward G+0.693$-$0.027 (shown in Fig. \ref{f:LTEspectrum}).}
\begin{tabular}{ccccccccccc}
\hline\hline
Frequency & Transition $^{(a)}$ & log \textit{I}& \textit{g}$\mathrm{_u}$ & $E$$\mathrm{_{LO}}$ & $E$$\mathrm{_{up}}$ & rms & $\int$ $T$$\mathrm{_A^*}$d$v$ & S/N $^{(b)}$ & Comments$^{(c)}$  \\ 
(GHz) & &  (nm$^2$ MHz) & & (cm$^{-1}$) &  (K ) & (mK) & (mK km s$^{-1}$) & & \\
\hline
40.5511839  & 3$_{1,3}–$2$_{1,2}$  & -6.0615  & 7 & 1.6 & 4.1 & 0.5 & 128.6 & 47.9 & Unblended   \\ 
40.5670170  & 3$_{0,3}$-2$_{0,2}$  & -6.0615  & 7 & 1.6 & 4.1 & 0.5$^{(d)}$ & 164.6 & 61.3 & Slightly blended: (CH$_{2}$OH)$_{2}$ \\
40.5674840  & 3$_{1,3}$-2$_{0,2}$  & -5.1434  & 7 & 1.6 & 4.1 &  &  &  & Slightly blended: (CH$_{2}$OH)$_{2}$ \\      
75.3102432  & 6$_{1,6}$-5$_{1,5}$*  & -5.1953  & 13 & 6.8 & 13.3 & 2.4 & 391.4 & 29.7 & Unblended$^\diamond$ \\ 
75.3102432  & 6$_{0,6}$-5$_{1,5}$*  & -4.2732  & 13 & 6.8 & 13.3 &  &  &  & Unblended$^\diamond$  \\ 
75.3102432  & 6$_{1,6}$-5$_{0,5}$*  & -4.2732  & 13 & 6.8 & 13.3 &  &  &  & Unblended$^\diamond$  \\ 
75.3102432  & 6$_{0,6}$-5$_{0,5}$*  & -5.1953  & 13 & 6.8 & 13.3 & &  &  & Unblended$^\diamond$   \\
75.3381091  & 5$_{1,4}$-4$_{2,3}$*  & -4.4608  & 11 & 6.0 & 9.7 & 2.4 & 228.4 & 17.3 & Unblended$^\diamond$  \\ 
75.3382086  & 5$_{2,4}$-4$_{2,3}$*  & -5.3723  & 11 & 6.0 & 9.7 &  &  &  & Unblended$^\diamond$  \\ 
75.3413737  & 5$_{1,4}$-4$_{1,3}$*  & -5.3722  & 11 & 6.0 & 9.7 &  &  &  & Unblended$^\diamond$   \\ 
75.3414732  & 5$_{2,4}$-4$_{1,3}$*  & -4.4607  & 11 & 6.0 & 9.7 &  &  &  & Unblended$^\diamond$   \\
86.8938108  & 7$_{0,7}$-6$_{0,6}$*  & -5.0043  & 15 & 6.0 & 17.4 & 1.2 & 68.5 & 10.3 & Unblended$^\diamond$     \\  
86.8938108  & 7$_{0,7}$-6$_{1,6}$*  & -4.0817  & 15 & 6.0 & 17.4 &  &  &  & Unblended$^\diamond$    \\  
86.8938108  & 7$_{1,7}$-6$_{0,6}$*  & -4.0817  & 15 & 6.0 & 17.4 &  &  &  & Unblended$^\diamond$     \\  
86.8938108  & 7$_{1,7}$-6$_{1,6}$*  & -5.0043  & 15 & 6.0 & 17.4 &  &  &  & Unblended$^\diamond$   \\
86.9222625  & 6$_{1,5}$-5$_{2,4}$*  & -4.2356  & 13 & 8.5 & 16.3 & 1.2 & 343.2 & 51.7 & Blended: HOCH$_{2}$CN and $g$-C$_{2}$H$_{5}$SH   \\ 
86.9222651  & 6$_{2,5}$-5$_{2,4}$*  & -5.1504  & 13 & 8.5 & 16.3 &  &  &  & Blended: HOCH$_{2}$CN and $g$-C$_{2}$H$_{5}$SH   \\ 
86.9223620  & 6$_{1,5}$-5$_{1,4}$*  & -5.1504  & 13 & 8.5 & 16.3 &  &  &  & Blended: HOCH$_{2}$CN and $g$-C$_{2}$H$_{5}$SH  \\ 
86.9223647  & 6$_{2,5}$-5$_{1,4}$*  & -4.2356  & 13 & 8.5 & 16.3 &  &  &  & Blended: HOCH$_{2}$CN and $g$-C$_{2}$H$_{5}$SH \\ 
88.7332608  & 4$_{4,1}$-3$_{3,0}$*  & -4.4485  & 9  & 4.7 & 10.9 & 1.6 & 86.1 & 9.7 & Unblended \\  
97.7784440  & 4$_{4,0}$-3$_{3,1}$*  & -4.5160  & 9  & 4.5 & 11.1 & 1.4 & 50.1 & 6.5 & Unblended \\  
98.4770930  & 8$_{0,8}$-7$_{0,7}$*  & -4.8401  & 17 & 12.2 & 22.1 & 1.4 & 94.7 & 12.2 & Blended: $t$-C$_{2}$H$_{3}$CHO \\    
98.4770930  & 8$_{0,8}$-7$_{1,7}$*  & -3.9172  & 17 & 12.2 & 22.1 &  &  &  & Blended: $t$-C$_{2}$H$_{3}$CHO   \\    
98.4770930  & 8$_{1,8}$-7$_{0,7}$*  & -3.9172  & 17 & 12.2 & 22.1 &  &  &  & Blended: $t$-C$_{2}$H$_{3}$CHO \\    
98.4770930  & 8$_{1,8}$-7$_{1,7}$*  & -4.8401  & 17 & 12.2 & 22.1 &  &  &  & Blended: $t$-C$_{2}$H$_{3}$CHO \\ 
98.5331999  & 6$_{2,4}$-5$_{3,3}$*  & -4.2241  & 13 &  9.9 & 18.8 & 1.4 & 174.2 & 22.5 & Blended: C$_{2}$H$_{5}$CN \\  
98.5336969 & 6$_{3,4}$-5$_{3,3}$*  & -5.1255  & 13 &  9.9 & 18.8 &  &   &  & Blended: C$_{2}$H$_{5}$CN \\  
98.5462274 & 6$_{2,4}$-5$_{2,3}$*  & -5.1254  & 13 &  9.9 & 18.8 & 1.4 & 72.5  & 9.4 & Blended: C$_{2}$H$_{5}$CN \\   
98.5467244 & 6$_{3,4}$-5$_{2,3}$*  & -4.2239  & 13 &  9.9 & 18.8 &  &   &  & Blended: C$_{2}$H$_{5}$CN \\ 
110.0599685  & 9$_{0,9}$-8$_{0,8}$*  & -4.6965  & 19 & 15.5 & 27.3 & 1.8 & 107.4 & 10.8 & Unblended$^\diamond$    \\   
110.0599685  & 9$_{0,9}$-8$_{1,8}$*  & -3.7733  & 19 & 15.5 & 27.3 &  &  &  & Unblended$^\diamond$   \\   
110.0599685  & 9$_{1,9}$-8$_{0,8}$*  & -3.7733  & 19 & 15.5 & 27.3 &  &  &  & Unblended$^\diamond$ \\   
110.0599685  & 9$_{1,9}$-8$_{1,8}$*  & -4.6965  & 19 & 15.5 & 27.3 &  &  &  & Unblended$^\diamond$   \\  
110.0880500 & 8$_{1,7}$-7$_{2,6}$*  & -3.8864  & 17 & 14.7 & 26.2 & 1.8 & 98.6 & 9.9 & Blended: CH$_{3}$OCH$_{3}$    \\  
110.0880500 & 8$_{2,7}$-7$_{2,6}$*  & -4.8046  & 17 & 14.7 & 26.2 &  &  &  & Blended: CH$_{3}$OCH$_{3}$    \\  
110.0880500 & 8$_{1,7}$-7$_{1,6}$*  & -4.8046  & 17 & 14.7 & 26.2 &  &  &  & Blended: CH$_{3}$OCH$_{3}$      \\  
110.0880500 & 8$_{2,7}$-7$_{1,6}$*  & -3.8864  & 17 & 14.7 & 26.2 &  &  &  & Blended: CH$_{3}$OCH$_{3}$    \\ 
\hline 
\end{tabular}
\label{tab:h2co3unblended}
\vspace*{-0.5ex}
\tablecomments{$^{(a)}$ The rotational energy levels are labelled using the conventional notation for asymmetric tops: $J_{K_{a},K_{c}}$, where $J$ denotes the angular momentum quantum number, and the $K_{a}$ and $K_{c}$ labels are projections of $J$ along the $a$ and $c$ principal axes. Lines that have been measured for the first time in the present astronomical dataset are marked with a * symbol. $^{(b)}$ The signal to noise (S/N) ratio is calculated from the integrated signal ($\int$ $T$$\mathrm{_A^*}$d$v$) and noise level $\sigma$ = rms $\times$ $\sqrt{\delta v \times \mathrm{FWHM}}$, where $\delta$$v$ is the velocity resolution of the spectra and the FWHM is fitted from the data. $^{(c)}$ We denote as “unblended" lines that are not contaminated by other species and use a $\diamond$ symbol for those that are (auto)blended with another transition of \textit{cis-trans} HOCOOH. $^{(d)}$ For those transitions that are partially or fully coalesced, we provide the integrated intensity and S/N ratio of the mean observed line rather than the values of each rotational transition, which is given only once for each group of transitions. The spectroscopic information was obtained from ctHOCOOH.cat, which is provided as Supporting Information.}
\end{table*}

We collect in Table \ref{tab:h2co3unblended} the most intense -according to the integrated S/N ratio- unblended or slightly blended transitions of \textit{cis-trans} HOCOOH detected toward G+0.693. The rest of the lines, which appear blended with brighter transitions from other species, show predicted intensities that are consistent with the observed spectra considering the contribution from all the species previously identified toward G+0.693. In Figure \ref{f:LTEspectrum} we depict the fitted line profiles of \textit{cis-trans} HOCOOH (red solid line) with integrated S/N ratio $>$ 6. Among them, we managed to detect up to four pairs of lines of \textit{cis-trans} HOCOOH, corresponding to different $K$$_a$ = 0,1 and 2 progressions, which reproduce almost perfectly the observations. Two of them are completely clean and their relative line intensities are well fitted (first and second panel of Fig. \ref{f:LTEspectrum}). While for the other two pairs (third and seventh panels), one line appears unblended and the other reproduces the observed spectra once the contribution from all other identified species is deemed (blue solid line). 

Then, we used all the above lines to carry out the corresponding LTE fit and to obtain the physical parameters of the emission of \textit{cis-trans} HOCOOH. We employed the \textsc{Autofit} tool within \textsc{Madcuba-Slim} \citep{martin2019}, which performs a nonlinear least-squares fitting of the simulated LTE spectra to the observed data. Note that we also considered the predicted emission from the already identified molecules in the same frequency region. The best-fitting LTE model for \textit{cis-trans} HOCOOH gives a molecular column density of $N$ = (6.4 $\pm$ 0.4) $\times$10$^{12}$ cm$^{-2}$, an excitation temperature of $T_{\rm ex}$ = 7.2 $\pm$ 0.6 K, a radial velocity of $v$$_{\rm LSR}$ = 69.8 $\pm$ 0.8 ~km~s$^{-1}$ and a full width half maximum of FWHM = 20 km s$^{-1}$ (fixed in the fit, according to the line width of the low-frequency $Q$-band lines). This value of the molecular column density is transcribed into a molecular abundance with respect to molecular hydrogen of $\sim$4.7 $\times$ 10$^{-11}$, adopting $N$(H$_{2}$) = 1.35$\times$10$^{23}$ cm$^{-2}$ from \citet{martin_tracing_2008}.

\begin{table*}
\centering
\caption{Derived physical parameters for HCOOH, CH$_3$COOH and HOCOOH toward the G+0.693-0.027 molecular cloud.}
\begin{tabular}{ c c c c c c c c  }
\hline
\hline
 Molecule & Formula & $N$   &  $T_{\rm ex}$ & $v$$_{\rm LSR}$ & FWHM  & Abundance$^a$ & Ref.$^b$  \\
 & & ($\times$10$^{14}$ cm$^{-2}$) & (K) & (km s$^{-1}$) & (km s$^{-1}$) & ($\times$10$^{-10}$) &   \\
\hline
\textit{Trans} formic acid & $t$-HCOOH  &  2.0 $\pm$ 0.4 & 10 $\pm$ 1 & 68 $\pm$ 2 & 22 $\pm$ 5 & 15 $\pm$ 4 &  (1)  \\
\textit{Cis} formic acid $^c$ & $c$-HCOOH  &  0.017 $\pm$ 0.002  & 10$^d$  & 69$^d$  & 17  $\pm$ 3 & 0.13  $\pm$ 0.02 &  (2)  \\
\hline
Acetic acid & CH$_3$COOH & 0.42 $\pm$ 0.02 &  17 $\pm$ 2 & 69.5 $\pm$ 0.5 & 21$^d$ & 3.1 $\pm$ 0.2 &  (2) \\
\hline
\textit{Cis-trans} carbonic acid & $ct$-HOCOOH & 0.064 $\pm$ 0.004 & 7.2 $\pm$ 0.6 & 69.8 $\pm$ 0.8 & 20$^d$ &  0.47 $\pm$ 0.03  &  (2) \\
\textit{Cis-cis} carbonic acid & $cc$-HOCOOH  & $\leq$ 1.6 & 7.2 & 69.8 & 20 & $\leq$ 12 &  (2) \\
\hline 
\end{tabular}
\label{tab:comparisonacids}
\vspace{0mm}
\vspace*{-0.5ex}
\tablecomments{$^a$ We adopted $N_{\rm H_2}$ = 1.35$\times$10$^{23}$ cm$^{-2}$, from \citet{martin_tracing_2008}. $^b$ References: (1) \citet{rodriguez-almeida2021a}; (2) This work; $^c$ For $c$-HCOOH we present a tentative detection. $^d$ Value fixed in the fit.}
\label{tab:g0693}
\end{table*}

\begin{figure}
\centerline{\resizebox{1\hsize}{!}{\includegraphics[angle=0]{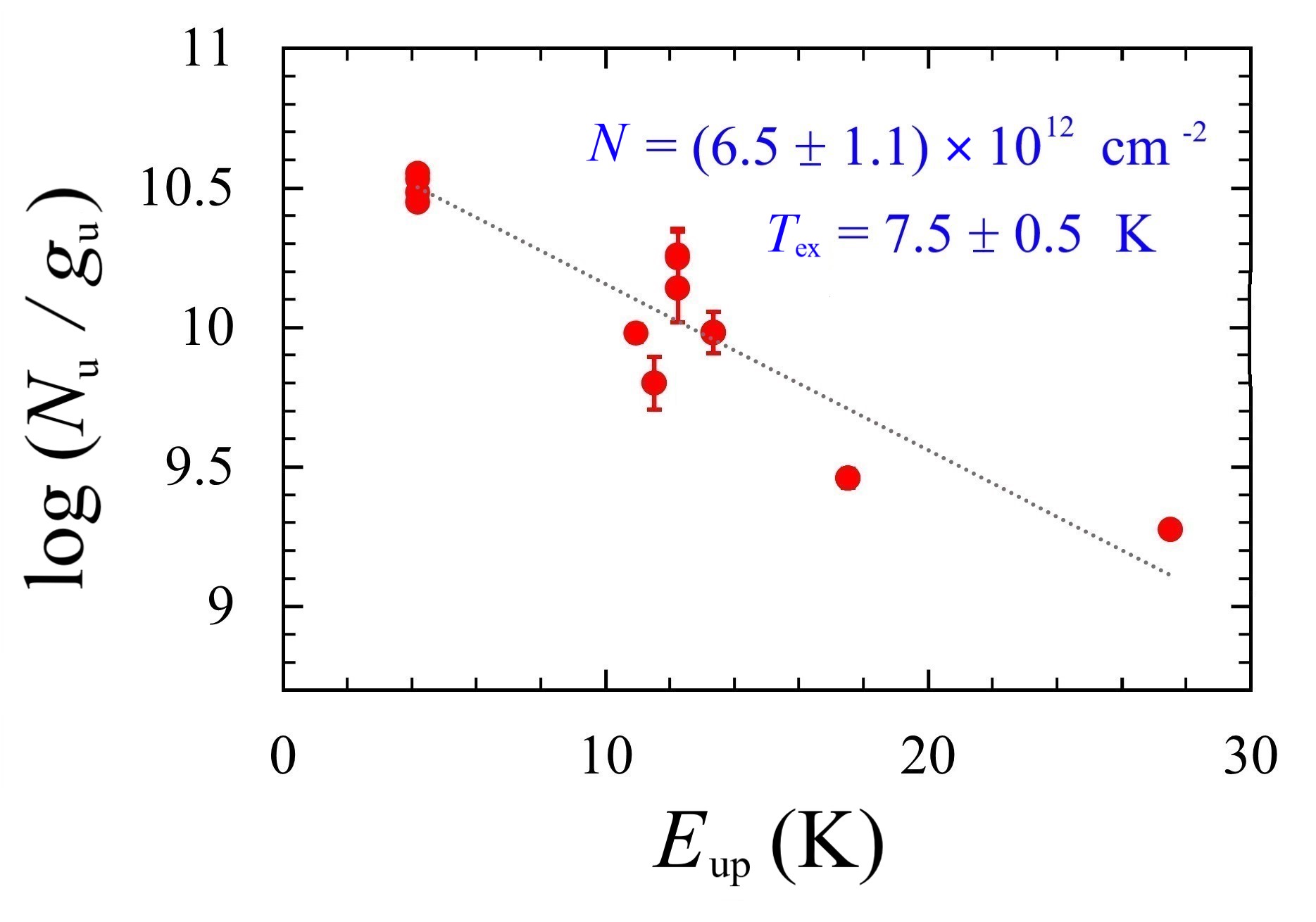}}}
\caption{Rotational diagram of \textit{cis-trans} HOCOOH using the unbleded transitions (depicted in red dots) reported in Table \ref{tab:h2co3unblended}. The gray dotted line corresponds to the best linear fit to the data points. The derived values for the molecular column density, $N$ and the excitation temperature, $T_{\rm ex}$, are shown in blue.}
\label{f:rotdiagram}
\end{figure}

We have also carried out a complementary population or rotational diagram analysis  \citep{goldsmith1999}, as implemented in \textsc{Madcuba,} using the velocity integrated intensity over the linewidth \citep{rivilla2021a}. Following this approach, we derived physical parameters for \textit{cis-trans} HOCOOH that are in perfect agreement with the \textsc{Autofit} analysis: $N$ = (6.5 $\pm$ 1.1) $\times$10$^{12}$ cm$^{-2}$, and $T_{\rm ex}$ = 7.5 $\pm$ 0.5 K. The results are shown in Fig. \ref{f:rotdiagram}.

Afterward, we searched for the global minimum in energy \textit{cis-cis} HOCOOH, but it was not detected in the current line survey due to its low dipole moment. To derive an upper limit to the column density, we adopted parameters obtained for the \textit{cis-trans} conformer (see Table \ref{tab:comparisonacids}) leaving the column density as the only free parameter. Then, we used the brightest transition according to the simulated LTE spectrum that is not contaminated due by other species. The brightest transition appears contaminated with the emission of C$_{2}$H$_{3}$CN and CH$_{3}$CHO. Thus, the chosen rotational transition for \textit{cis-cis} HOCOOH is the 3$_{1,3}$-2$_{0,2}$ (located at 40.7460 GHz), from which we obtained a 3$\sigma$ upper limit to the column density of $N$ $\leq$ 1.6$\times$10$^{14}$ cm$^{-2}$ that does not produce any overly bright features at other frequencies. In terms of the molecular abundance with respect to H$_{2}$, the above value is translated into an upper limit of $\leq$1.2$\times$10$^{-9}$, which further constrains the abundance of \textit{cis-cis} HOCOOH in the ISM. The results of the LTE fits are listed in Table \ref{tab:comparisonacids}.

\subsection{Detection of CH$_3$COOH and tentative detection of c-HCOOH} 
\label{subsec:detectionacetic}

Acetic acid, CH$_3$COOH, has been previously searched for toward G+0.693 (an upper limit to the column density was reported in \citealt{requena-torres_organic_2006}). Herein, we performed an accompanying search using the current astronomical dataset, and CH$_3$COOH is now unequivocally detected (see Fig. \ref{f:LTECH3COOH}). We present the complete LTE analysis in the Appendix \ref{AnalysisCH3COOH} (see Table \ref{AnalysisCH3COOH}), and the derived physical parameters are collected in Table \ref{tab:comparisonacids} along with the physical parameters of HCOOH and HOCOOH. 

Regarding formic acid, HCOOH, the \textit{trans} conformer is commonly found in regions with some degree of chemical complexity, including G+0.693 \citep{rodriguez-almeida2021a,garciadelaconcepcion2022}. However, the less stable \textit{cis} conformer has only been found toward sources displaying specific physical and chemical properties, such as the Orion Bar photo-dissociation region  \citep{Cuadrado:2016hp} and the cold molecular clouds L483 \citep{agundez2019} and B5 \citep{taquet2017}. In this work, we managed to tentatively detect \textit{cis}-HCOOH using the new ultra-deep observations of G+0.693. We present the corresponding LTE analysis in the Appendix \ref{AnalysisHCOOH} (see Table \ref{tab:LTEHCOOH} and Figure \ref{f:LTEHCOOH}), and we list the derived physical parameters in the second row of Table \ref{tab:comparisonacids}.

\section{Discussion} 
\label{sec:disc}

\subsection{Conformational isomerism of HOCOOH}
\label{subsec:othermolecules}

On the basis of the derived molecular abundances (see Table \ref{tab:comparisonacids}), we obtain a \textit{cis-cis}/\textit{cis-trans} HOCOOH abundance ratio of $\leq$25, which is relatively lower than the \textit{trans/cis} upper limit ratio obtained for the related HCOOH in G+0.693 ($\sim$ 117) considering the tentative detection in G+0.693 presented here, and also in the high-mass star-forming region G31.41+0.31 ($\geq$ 90; \citealt{garciadelaconcepcion2022}). Nevertheless, this fact can be easily rationalized in terms of the different relative electronic energies between the two low-lying conformers of HCOOH  ($\Delta$$E$(\textit{cis}/\textit{trans}) = 4.04 kcal mol$^{-1}$; \citealt{garciadelaconcepcion2022}) and HOCOOH ($\Delta$$E$(\textit{cis-trans}/\textit{cis-cis}) =1.71 kcal mol$^{-1}$; \citealt{Mori2009}), respectively. A similar yet more pronounced behavior was observed between HCOOH and HC(O)SH, since the relative electronic energy between the two conformers of the S-bearing acid is only 0.68 kcal mol$^{-1}$, which is translated in a much lower \textit{trans/cis} abundance ratio of $\sim$3.7 \citep{garciadelaconcepcion2022}. 

Similarly to HCOOH, HOCOOH can also exhibit \textit{trans/cis} rotational or conformational isomerism \citep{Petterson2002,Macoas2005,Tsuge2015,garciadelaconcepcion2022}, and the stability of each conformer is driven by the formation of stabilizing intramolecular hydrogen bonds. These type of isomerization processes have already been proven to be feasible under ISM conditions thanks to multi-dimensional ground-state quantum tunnelling effects \citep{garciadelaconcepcion2022}. Therefore, given that HOCOOH can also undergo quantum tunnelling \citep{Wagner2016}, we suggest that the \textit{cis-cis}/\textit{cis-trans} ratio observed in the ISM may be explained in terms of the relative stability of the conformers, which is consistent with the low abundance reported for the high-energy \textit{cis-trans} HOCOOH. For several stereoisomers detected toward G+0.693 we have observed that the derived relative isomeric match that predicted under thermodynamic equilibrium conditions, further supporting the above statement, highlighting: $Z$ and $E$ isomers of cyanomethanimine (HNCHCN), whose relative ratio was rationalized theoretically under ISM conditions also based on quantum tunnelling effects \citep{garciadelaconcepcion2021}, and the $Ga$ and $Aa$ conformers of $n$-propanol ($n$-C$_3$H$_7$OH; \citealt{jimenez-serra2022}).



Furthermore, although we have shown that the detection of \textit{cis-cis} HOCOOH by means of its rotational spectra is at the very least a challenging task, we can now shed light on the total abundance of the molecule based on the detection of the higher-in-energy \textit{cis-trans} form, which exhibits a sizable dipole moment. This fact shall guide the astronomical community to perform future searches for referable systems toward sources exhibiting high kinetic temperatures ($T$$_{\rm kin}$), such as Galactic Center molecular clouds (i.e. G+0.693), hot cores and corinos, which can populate higher-in-energy species efficiently, if the energy difference is not exorbitant and assuming a thermodynamic population.

\begin{table*}
\centering
\tabcolsep 3.0pt
\caption{Abundances and ratio between formic and acetic acid toward different astronomical environments.}
\begin{tabular}{ c  c c c c c c }
\hline
\hline
Source & $N$($t$-HCOOH)  &  $N$(CH$_3$COOH)  & $N$(C$_2$O$_2$H$_4$)$^c$ & $\mathrm{\frac{t-HCOOH}{CH_3COOH}}$ & $\mathrm{\frac{t-HCOOH}{C_2O_2H_4}}$ & Ref.$^d$  \\
  & (cm$^{-2}$) & (cm$^{-2}$) & (cm$^{-2}$) &   &    \\
\hline
G+0.693  &  (2.0 $\pm$ 0.4) $\times$10$^{14}$   & (4.5 $\pm$ 0.2) $\times$10$^{13}$ & (8.4 $\pm$ 0.6) $\times$10$^{14}$ & 4.4  & 0.24 & (1)  \\
Sgr B2(N) & (1.5 $\pm$ 0.2) $\times$10$^{16}$  & (1.1 $\pm$ 0.1) $\times$10$^{16}$  & (4.5 $\pm$ 0.5) $\times$10$^{17}$  &  1.4 &  0.03 &  (2) \\
67P/C-G & 0.013 $\pm$ 0.008 $^a$ & -  & 0.034 $\pm$ 0.002  & - & 0.4 & (3) \\
Ryugu (A0106) & (9.47  $\pm$ 10) $\times$10$^{3}$ $^b$ & (5.7 $\pm$ 1.5) $\times$10$^{3}$ & - & 1.7 & - &  (4) \\
\hline 
\end{tabular}
\label{tab:comparisonsources}
\vspace{0mm}
\vspace*{-0.5ex}
\tablecomments{$^a$ In 67P/C-G the derived abundance is that relative to H$_2$O (per cent). $^b$ In the hot-water extract (A0106) of Ryugu, the abundance is reported in nmol g$^{-1}$. $^c$ The relative abundance derived in 67P/C-G is referred to all the C$_2$O$_2$H$_4$ isomers and apart from CH$_3$COOH it also includes the isomers methyl formate (HCOOCH$_3$), glycolaldehyde (CH$_2$OHCHO) and ethenediol, (CHOH)$_2$, (only detected toward G+0.693). $^d$ References: (1) \citet{rodriguez-almeida2021a,rivilla2022a} and this work; (2) \citet{belloche_complex_2013}; (43) \citet{Drozdovskaya2019}; (4) \citet{Naraoka2023}.}
\label{tab:othersources}
\end{table*}

\subsection{Relative abundance of carboxylic acids in space}
\label{subsec:othermolecules}

To put HOCOOH in a broader astrochemical context, we can compare its abundance in G+0.693 with the abundance of the only two detected carboxilic acids in the ISM so far, also detected toward G+0.693: HCOOH and CH$_3$COOH. 

As shown in Table \ref{tab:comparisonacids}, we obtained a CH$_3$COOH/$ct$-HOCOOH ratio of $\sim$7, showing that CH$_3$COOH is not significantly more abundant than the high-energy form of HOCOOH toward G+0.693. Moreover, we found that \textit{cis-trans} HOCOOH is $\sim$31 times less abundant than $t$-HCOOH and it is, at the same time, $\sim$4 times more abundant than $c$-HCOOH toward this astronomical source. Nevertheless, if we consider the upper limit derived for the \textit{cis-cis} form of HOCOOH and the aforementioned conformational isomerism, its abundance is expected to be of the same order as that of \textit{t}-HCOOH. Thus, carbonic acid possibly emerges as an ubiquitous and abundant O-bearing COM in the ISM although it went unnoticed so far.


At this point, the detection of both HCOOH and CH$_3$COOH toward G+0.693 allows us to provide a rough comparison with the concentration of these acids in the envelope of the neighbour star-forming region Sgr B2(N) \citep{belloche_complex_2013} as well as in asteroids (in the Ryugu material; \citealt{Naraoka2023}) and comets (67P/Churyumov-Gerasimenko, 67P/C-G; \citealt{altwegg2016,Drozdovskaya2019}). The results are summarized in Table \ref{tab:comparisonsources} and Figure \ref{f:ratioref} (Appendix \ref{Figratio}). As it can be seen, the HCOOH/CH$_3$COOH ratio derived in the Ryugu material ($\sim$1.7; \citealt{Naraoka2023}) is consistent with the one obtained through radio astronomical observations in both G+0.693 ($\sim$4.4) and the envelope of Sgr B2(N) ($\sim$1.4), within an factor of three. In comparison, a larger amount of HCOOH is available in the G+0.693 molecular cloud, which will most likely react to form more complex derivatives at later stages. For CH$_3$COOH, the mass spectrometric measurements carried out with the ROSINA instrument \citep{Drozdovskaya2019} were not able to differentiate between acetic acid and their isomers: methyl formate (HCOOCH$_3$), glycolaldehyde (CH$_2$OHCHO) and ethenediol, (CHOH)$_2$. Thus, we need to include these isomers (C$_2$O$_2$H$_4$) to properly analyze the abundance of CH$_3$COOH in the volatile cometary material of 67P/C-G, which exhibits a similar HCOOH/C$_2$O$_2$H$_4$ ratio ($\sim$0.4) compared with that obtained in G+0.693 ($\sim$0.24). Meanwhile, in Sgr B2(N) a ratio one order of magnitude lower is observed, due to the large abundance of methyl formate toward this source. Our results, therefore, suggest a relationship between the relative molecular abundances of carboxylic acids in the ISM and that found in minor bodies of the Solar System, suggesting they survive the star formation process \citep{rivilla2021a}.

The fact that the identification of HOCOOH in extraterrestrial bodies (e.g., asteroids and comets) remains elusive is not surprising. In this context, the debate whether HOCOOH could persist or not long enough to enable its spectroscopic characterization remained open for many years, since this elusive molecule rapidly decomposes into CO$_2$ and H$_2$O \citep{Ghoshal2015} or upon deprotonatation in solution. This affects significantly its tractability in aqueous solution -the presence of just two catalytically active water molecules leads to decomposition \citep{Loerting2000}- but also in the presence of other organic molecules \citep{Ghoshal2015}. In this context, aqueous alteration might have happened in the Ryugu samples but it is not likely to have occurred to an important extent in comet 67P/C-G, whose composition is expected to be more pristine. However, gaseous HOCOOH presents a rather surprising kinetic stability in the absence of water \citep{Loerting2000}. Fortunately, it was identified at last in gas-phase isolation conditions by \citet{Mori2009,Mori2011}, which guided us to achieve the present interstellar discovery. In addition, although to our knowledge HOCOOH has not been searched for in the volatile cometary material of 67P/C-G, its possible identification might have to face a problem of contamination of its mass peak with ethylene glycol, (CH$_2$OH)$_2$ \citep{rubin2019}, which shares the same molecular mass of HOCOOH (63.02 u). Thus, it could indeed be present in the extraterrestrial material of asteroids and comets but it yet remains unidentified.

\subsection{Interstellar formation of HOCOOH}
\label{subsec:formation}

The formation of HOCOOH in the ISM has been the subject of a plethora of recent experimental and theoretical studies (e.g., \citealt{Gerakines2000,Zheng2007,Baltrusaitis2010,Oba2010,Ioppolo2021}).
\cite{Oba2010} investigated the formation of HOCOOH on dust-grain surfaces from the reaction between CO and the radical species OH to form both \textit{cis} and \textit{trans} HOCO at low temperatures (\citealt{Lester2001,Noble2011,Nguyen2012,Ruad2015,Tachikawa2016}; reactions \ref{eq2},\ref{eq1}). 

\begin{equation}
\mathrm{CO} + \mathrm{OH} \rightarrow \rm \textit{trans-}\mathrm{HOCO} 
\label{eq2}
\end{equation}
\begin{equation}
\mathrm{CO} + \mathrm{OH} \rightarrow \mathrm{\textit{cis-}HOCO}
\label{eq1}
\end{equation}
\begin{equation}
\mathrm{\textit{trans-}HOCO} + \mathrm{OH} \rightarrow \mathrm{HOCOOH} 
\label{eq4}
\end{equation}
\begin{equation}
\mathrm{\textit{cis-}HOCO} + \mathrm{OH} \rightarrow \mathrm{HOCOOH} 
\label{eq3}
\end{equation}

Once HOCO is formed, it may fall apart to yield H + CO$_2$ or further react with additional OH radicals yielding HOCOOH (\ref{eq4},\ref{eq3}). This surface pathway has already been suggested as one of the key plausible routes on the formation of HOCOOH in cold dense molecular clouds \citep{Ioppolo2021}. It again emphasizes the relevance of radical addition reactions, in particular those involving reactions between the OH radical and neutral species in interstellar surface chemistry (e.g., \citealt{Molpeceres2022}). In addition, although the most stable form of the HOCO radical, \textit{trans}-HOCO, remains undetected toward G+0.693, based on the derived integrated intensity 3$\sigma$ upper limit to its column density of N $\leq$ 1.2 $\times$10$^{12}$ cm$^{-2}$ (using entry 45517 of the CDMS catalog; \citealt{Muller2005}), we obtain a \textit{ct}-HOCOOH/\textit{t}-HOCO abundance ratio of $\sim$5.3. This ratio could be rationalized based on the low abundance expected for such intermediate and highly reactive species in the suggested formation pathways, which may perhaps be formed but will rapidly react to form more complex products.

Moreover, HOCOOH has been hinted as the main reaction product of the energetic processing of H$_2$O/CO$_2$ icy mixtures (in different proportions), highlighting the electron-induced formation route proposed by \cite{Zheng2007}, which is initiated by reaction (\ref{eq5}) and (\ref{eq6}), followed by the radical association with OH (see reaction \ref{eq3} above):

\begin{equation}
\mathrm{H_2O} + \mathrm{e^-} \rightarrow \mathrm{H} + \mathrm{OH} + \mathrm{e^-}
\label{eq5}
\end{equation}
\begin{equation}
\mathrm{H} + \mathrm{CO_2} \rightarrow \mathrm{\textit{cis-}HOCO} 
\label{eq6}
\end{equation}

Alternatively, the production of HOCOOH may occur through atomic-H abstraction from HCOOH \citep{Markmeyer2019,Molpeceres2022a}, which is efficiently formed in the ISM \citep{Qasim2019}, followed by radical addition (i.e. HOCO + OH). Therefore, sources that are reservoirs of large amounts of HCOOH are excellent targets for confirming the detection of HOCOOH. 

To our knowledge, apart from the direct CO$_2$ + H$_2$O reaction that shows high activation barriers \citep{Baltrusaitis2010}, gas-phase formation routes of HOCOOH remain unexplored. For instance, we can suggest ion-molecule reactions between positive ions, such as the detected HOCO$^+$ \citep{Thaddeus1981,Majumdar2018}, which is also present in G+0.693 \citep{armijos-abendano2015}, or even HCOOH$^+$, and neutral molecules.

Hence, although HOCOOH is most likely formed via OH radical addition to HOCO on the surface of dust-grains, further theoretical effort is needed to build complete chemical reaction networks to understand the presence of HOCOOH in space and to decipher which of the above formation pathways prevail over the rest. Finally, HOCOOH represents the first molecule with more than two O atoms detected so far in the ISM. Thus, the current investigation expands our knowledge on O-bearing COMs and opens the window for the detection of other interstellar molecules containing multiple oxygen atoms.

\section{summary and conclusions} \label{sec:con}

Modern radio astronomical facilities and laboratory work are steadily pushing the limits of the chemical complexity found in the interstellar medium. However, for certain families of molecules (i.e., carboxylic acids) the count of identified interstellar species has remained unchanged for more than two decades. In this context, \textit{cis-trans} HOCOOH has been detected at last in the ISM toward the G+0.693-0.027 Galactic Center molecular cloud, and stands as the first interstellar molecule containing three O atoms. We derive a molecular column density for this conformer of $N$ = (6.4 $\pm$ 0.4) $\times$ 10$^{12}$ cm$^{-2}$, which is translated into a molecular abundance with respect to molecular H$_2$ of 4.7 $\times$ 10$^{-11}$. Although the available laboratory measurements were limited to 65 GHz, we used our spectral line survey as a “conventional” laboratory spectrum and managed to detect several clear and unblended spectroscopic features directly in the radio astronomical data (between 75-120 GHz). This fact further allowed us to slightly improve the available set of rotational spectroscopic constants of \textit{cis-trans} HOCOOH. We also report the nondetection of the more stable \textit{cis-cis} HOCOOH, showing an upper limit to the molecular abundance with respect to H$_2$ of $\leq$ 1.2 $\times$10$^{-9}$, due to the huge impact that its low dipole moment has on the overall detectability of the conformer. Nevertheless, based on the derived upper limit, it is still likely that \textit{cis-cis} HOCOOH is rather abundant. We suggest that it may be efficiently formed under interstellar conditions, most likely via -OH radical addition of HOCO on the surface of dust grains. Moreover, we obtained a \textit{cis-cis}/\textit{cis-trans} ratio $\leq$ 25, in good agreement with the \textit{trans}/\textit{cis} ratio observed for the related HCOOH in G+0.693 based on the larger energetic difference between both HCOOH conformers. We found that \textit{cis-trans} carbonic acid is $\sim$31 and 7 times less abundant than $t$-HCOOH and CH$_3$COOH, respectively, toward this astronomical source, but it is also $\sim$4 times more abundant than $c$-HCOOH. In addition, our comparison between the abundance of HCOOH and CH$_3$COOH in different astronomical environments, including star-forming regions, asteroids and comets, enabled us to establish an overall good correlation between the relative molecular abundance of carboxylic acids.


The interstellar discovery of carbonic acid (HOCOOH) presented in this work provides relevant insight into the actual degree of chemical complexity of the ISM, and will bear significant implications to unravel the role of HOCOOH within interstellar carbon and oxygen chemistry, based on the large amount of HOCOOH that might be lurking in space. This study will also pave the route to perform new harmonized observational, theoretical and laboratory investigations targeting astronomical candidates that exhibit an extremely low dipole moment (e.g., \textit{cis-cis} HOCOOH) based on the detection of other moderately higher-in-energy conformers with sizable dipole moments. Hence, we now open the door to achieve indirect interstellar identifications of conformers that remained undetectable to radioastronomy, specially toward high-$T$$_{\rm kin}$ sources such as Galactic Center molecular clouds (i.e. G+0.693), or hot cores and corinos, which are capable of populating these high-energy species efficiently.




\software{1) Madrid Data Cube Analysis (\textsc{Madcuba}) on ImageJ is a software developed at the Center of Astrobiology (CAB) in Madrid; \url{http://cab.inta-csic.es/madcuba/}; \citet{martin2019}; version 9.3.10 (04/05/2023).
2). \textsc{Gildas} package: \url{https://www.iram.fr/IRAMFR/GILDAS} 3) Python: \url{https://www.python.org}. 3) Pickett's SPFIT/SPCAT program suite; \url{https://spec.jpl.nasa.gov/}; \citet{Pickett:1991cv}.}

\begin{acknowledgments}
We are grateful to the IRAM 30m and Yebes 40m telescopes staff for their help during the different observing runs. The 40m radio telescope at Yebes Observatory is operated by the Spanish Geographic Institute (IGN, Ministerio de Transportes, Movilidad y Agenda Urbana). IRAM is supported by INSU/CNRS (France), MPG (Germany) and IGN (Spain). M.S.N. thanks the financial funding from the European Union - NextGenerationEU, Ministerio de Universidades and the University of Valladolid under a postdoctoral Margarita Salas Grant, as well as financial support from the Spanish Ministerio de Ciencia e Innovaci{\'o}n (PID2020-117742GB-I00). V.M.R., I.J.-S., J.M.-P., L.C, A.M., and A.M.-H. acknowledge financial support from projects number RYC2020-029387-I and PID2019-105552RB-C41 funded by the Spanish Ministry of Science and Innovation/State Agency of Research MCIN/AEI/10.13039/501100011033. A.M. has received support from the Spanish project number MDM-2017-0737-19-2 and grant PRE2019-091471, funded by MCIN/AEI/10.13039/501100011033 and by “ESF Investing in your future. A.M.-H acknowledges funds from Grant MDM-2017-0737 Unidad de Excelencia “Mar{\'i}a de Maeztu" Centro de Astrobiolog{\'i}a (CAB, INTA-CSIC). D.S.A acknowledges the funds provided by the Consejo Superior de Investigaciones Cient{\'i}ficas (CSIC) and the Centro de Astrobiolog{\'i}a (CAB) through the project 20225AT015 (Proyectos intramurales especiales del CSIC).” P.dV. and B.T. thank the support 
from the Spanish Ministerio de Ciencia e Innovaci\'on (MICIU) through project PID2019-107115GB-C21. B.T. also thanks the Spanish MICIU for funding support from grant PID2019-106235GB-I00.
 
\end{acknowledgments}




\bibliography{rivilla,bibliography}{}

\begin{thebibliography}{}
\expandafter\ifx\csname natexlab\endcsname\relax\def\natexlab#1{#1}\fi
\providecommand{\url}[1]{\href{#1}{#1}}
\providecommand{\dodoi}[1]{doi:~\href{http://doi.org/#1}{\nolinkurl{#1}}}
\providecommand{\doeprint}[1]{\href{http://ascl.net/#1}{\nolinkurl{http://ascl.net/#1}}}
\providecommand{\doarXiv}[1]{\href{https://arxiv.org/abs/#1}{\nolinkurl{https://arxiv.org/abs/#1}}}

\bibitem[{{Adamczyk} {et~al.}(2009){Adamczyk}, {Pr{\'e}mont-Schwarz}, {Pines},
  {Pines}, \& {Nibbering}}]{Adamczyk2009}
{Adamczyk}, K., {Pr{\'e}mont-Schwarz}, M., {Pines}, D., {Pines}, E., \&
  {Nibbering}, E. T.~J. 2009, Science, 326, 1690,
  \dodoi{10.1126/science.1180060}

\bibitem[{{Ag{\'u}ndez} {et~al.}(2019){Ag{\'u}ndez}, {Marcelino}, {Cernicharo},
  {Roueff}, \& {Tafalla}}]{agundez2019}
{Ag{\'u}ndez}, M., {Marcelino}, N., {Cernicharo}, J., {Roueff}, E., \&
  {Tafalla}, M. 2019, \aap, 625, A147, \dodoi{10.1051/0004-6361/201935164}

\bibitem[{{Alonso} {et~al.}(2015){Alonso}, {Kolesnikov{\'a}}, {Pe{\~n}a},
  {Shipman}, {Tercero}, {Cernicharo}, \& {Alonso}}]{Alonso2015A}
{Alonso}, E.~R., {Kolesnikov{\'a}}, L., {Pe{\~n}a}, I., {et~al.} 2015, Journal
  of Molecular Spectroscopy, 316, 84, \dodoi{10.1016/j.jms.2015.08.002}

\bibitem[{Altwegg {et~al.}(2016)Altwegg, Balsiger, Bar-Nun, Berthelier, Bieler,
  Bochsler, Briois, Calmonte, Combi, Cottin, De~Keyser, Dhooghe, Fiethe,
  Fuselier, Gasc, Gombosi, Hansen, Haessig, Ja~ckel, Kopp, Korth, Le~Roy, Mall,
  Marty, Mousis, Owen, Reme, Rubin, Semon, Tzou, Hunter~Waite, \&
  Wurz}]{Altwegg:2016ck}
Altwegg, K., Balsiger, H., Bar-Nun, A., {et~al.} 2016, Science Advances, 2,
  e1600285

\bibitem[{{Altwegg} {et~al.}(2016){Altwegg}, {Balsiger}, {Bar-Nun},
  {Berthelier}, {Bieler}, {Bochsler}, {Briois}, {Calmonte}, {Combi}, {Cottin},
  {De Keyser}, {Dhooghe}, {Fiethe}, {Fuselier}, {Gasc}, {Gombosi}, {Hansen},
  {Haessig}, {Ja ckel}, {Kopp}, {Korth}, {Le Roy}, {Mall}, {Marty}, {Mousis},
  {Owen}, {Reme}, {Rubin}, {Semon}, {Tzou}, {Waite}, \& {Wurz}}]{altwegg2016}
{Altwegg}, K., {Balsiger}, H., {Bar-Nun}, A., {et~al.} 2016, Science Advances,
  2, e1600285, \dodoi{10.1126/sciadv.1600285}

\bibitem[{{Armijos-Abenda{\~n}o} {et~al.}(2015){Armijos-Abenda{\~n}o},
  {Mart{\'\i}n-Pintado}, {Requena-Torres}, {Mart{\'\i}n}, \&
  {Rodr{\'\i}guez-Franco}}]{armijos-abendano2015}
{Armijos-Abenda{\~n}o}, J., {Mart{\'\i}n-Pintado}, J., {Requena-Torres}, M.~A.,
  {Mart{\'\i}n}, S., \& {Rodr{\'\i}guez-Franco}, A. 2015, \mnras, 446, 3842,
  \dodoi{10.1093/mnras/stu2271}

\bibitem[{{Baltrusaitis} \& {Grassian}(2010)}]{Baltrusaitis2010}
{Baltrusaitis}, J., \& {Grassian}, V.~H. 2010, Journal of Physical Chemistry A,
  114, 2350, \dodoi{10.1021/jp9097809}

\bibitem[{Belloche {et~al.}(2013)Belloche, M\"uller, Menten, Schilke, \&
  Comito}]{belloche_complex_2013}
Belloche, A., M\"uller, H. S.~P., Menten, K.~M., Schilke, P., \& Comito, C.
  2013, Astronomy and Astrophysics, 559, A47,
  \dodoi{10.1051/0004-6361/201321096}

\bibitem[{{Bennett} {et~al.}(2014){Bennett}, {Ennis}, \&
  {Kaiser}}]{Bennett2014}
{Bennett}, C.~J., {Ennis}, C.~P., \& {Kaiser}, R.~I. 2014, \apj, 794, 57,
  \dodoi{10.1088/0004-637X/794/1/57}

\bibitem[{{Blom} \& {Bauder}(1981)}]{Blom1981}
{Blom}, C.~E., \& {Bauder}, A. 1981, Chemical Physics Letters, 82, 492,
  \dodoi{10.1016/0009-2614(81)85426-7}

\bibitem[{{Caldeira} \& {Wickett}(2003)}]{Caldeira2003}
{Caldeira}, K., \& {Wickett}, E. M.~E. 2003, Nature, 425, 265,
  \dodoi{10.1038/425365a}

\bibitem[{{Combes} {et~al.}(1996){Combes}, {Q-Rieu}, \&
  {Wlodarczak}}]{combes1996}
{Combes}, F., {Q-Rieu}, N., \& {Wlodarczak}, G. 1996, \aap, 308, 618

\bibitem[{{Cooper} {et~al.}(1992){Cooper}, {Onwo}, \& {Cronin}}]{Cooper1992}
{Cooper}, G.~W., {Onwo}, W.~M., \& {Cronin}, J.~R. 1992, \gca, 56, 4109,
  \dodoi{10.1016/0016-7037(92)90023-C}

\bibitem[{{Cuadrado} {et~al.}(2016){Cuadrado}, {Goicoechea}, {Roncero},
  {Aguado}, {Tercero}, \& {Cernicharo}}]{Cuadrado:2016hp}
{Cuadrado}, S., {Goicoechea}, J.~R., {Roncero}, O., {et~al.} 2016, \aap, 596,
  L1, \dodoi{10.1051/0004-6361/201629913}

\bibitem[{{Cunningham} {et~al.}(2007){Cunningham}, {Jones}, {Godfrey}, {Cragg},
  {Bains}, {Burton}, {Calisse}, {Crighton}, {Curran}, {Davis}, {Dempsey},
  {Fulton}, {Hidas}, {Hill}, {Kedziora-Chudczer}, {Minier}, {Pracy}, {Purcell},
  {Shobbrook}, \& {Travouillon}}]{Cunningham07}
{Cunningham}, M.~R., {Jones}, P.~A., {Godfrey}, P.~D., {et~al.} 2007, \mnras,
  376, 1201

\bibitem[{{Delitsky} \& {Lane}(1998)}]{Delitsky1998}
{Delitsky}, M.~L., \& {Lane}, A.~L. 1998, \jgr, 103, 31391,
  \dodoi{10.1029/1998JE900020}

\bibitem[{{Delitsky} {et~al.}(2017){Delitsky}, {Paige}, {Siegler}, {Harju},
  {Schriver}, {Johnson}, \& {Travnicek}}]{Delitsky2017}
{Delitsky}, M.~L., {Paige}, D.~A., {Siegler}, M.~A., {et~al.} 2017, \icarus,
  281, 19, \dodoi{10.1016/j.icarus.2016.08.006}

\bibitem[{{Drozdovskaya}(2019)}]{Drozdovskaya2019}
{Drozdovskaya}, M.~N. 2019, in From Stars to Planets II - Connecting our
  understanding of star and planet formation, 5

\bibitem[{{Ehrenfreund} {et~al.}(2001){Ehrenfreund}, {Bernstein}, {Dworkin},
  {Sandford}, \& {Allamandola}}]{Ehrenfreund2001}
{Ehrenfreund}, P., {Bernstein}, M.~P., {Dworkin}, J.~P., {Sandford}, S.~A., \&
  {Allamandola}, L.~J. 2001, \apjl, 550, L95, \dodoi{10.1086/319491}

\bibitem[{{Garc{\'\i}a de la Concepci{\'o}n} {et~al.}(2021){Garc{\'\i}a de la
  Concepci{\'o}n}, {Jim{\'e}nez-Serra}, {Carlos Corchado}, {Rivilla}, \&
  {Mart{\'\i}n-Pintado}}]{garciadelaconcepcion2021}
{Garc{\'\i}a de la Concepci{\'o}n}, J., {Jim{\'e}nez-Serra}, I., {Carlos
  Corchado}, J., {Rivilla}, V.~M., \& {Mart{\'\i}n-Pintado}, J. 2021, \apjl,
  912, L6.
\newblock \doarXiv{2104.07913}

\bibitem[{{Garc{\'\i}a de la Concepci{\'o}n} {et~al.}(2022){Garc{\'\i}a de la
  Concepci{\'o}n}, {Colzi}, {Jim{\'e}nez-Serra}, {Molpeceres}, {Corchado},
  {Rivilla}, {Mart{\'\i}n-Pintado}, {Beltr{\'a}n}, \&
  {Mininni}}]{garciadelaconcepcion2022}
{Garc{\'\i}a de la Concepci{\'o}n}, J., {Colzi}, L., {Jim{\'e}nez-Serra}, I.,
  {et~al.} 2022, \aap, 658, A150, \dodoi{10.1051/0004-6361/202142287}

\bibitem[{{Georgiou} \& {Deamer}(2014)}]{Georgiou2014}
{Georgiou}, C.~D., \& {Deamer}, D.~W. 2014, Astrobiology, 14, 541,
  \dodoi{10.1089/ast.2013.1134}

\bibitem[{{Gerakines} {et~al.}(2000){Gerakines}, {Moore}, \&
  {Hudson}}]{Gerakines2000}
{Gerakines}, P.~A., {Moore}, M.~H., \& {Hudson}, R.~L. 2000, \aap, 357, 793

\bibitem[{Ghoshal \& Hazra(2015)}]{Ghoshal2015}
Ghoshal, S., \& Hazra, M.~K. 2015, RSC Adv., 5, 17623,
  \dodoi{10.1039/C4RA13233E}

\bibitem[{{Glavin} {et~al.}(2010){Glavin}, {Callahan}, {Dworkin}, \&
  {Elsila}}]{Glavin2010}
{Glavin}, D.~P., {Callahan}, M.~P., {Dworkin}, J.~P., \& {Elsila}, J.~E. 2010,
  \maps, 45, 1948, \dodoi{10.1111/j.1945-5100.2010.01132.x}

\bibitem[{{Goldsmith} \& {Langer}(1999)}]{goldsmith1999}
{Goldsmith}, P.~F., \& {Langer}, W.~D. 1999, \apj, 517, 209

\bibitem[{Gordy \& Cook(1984)}]{Gordy:1984uy}
Gordy, W., \& Cook, R.~L. 1984, {Microwave Molecular Spectra}, 3rd edn. (New
  York: Wiley)

\bibitem[{{Hage} {et~al.}(1998){Hage}, {Liedl}, {Hallbrucker}, \&
  {Mayer}}]{Hage1998}
{Hage}, W., {Liedl}, K.~R., {Hallbrucker}, A., \& {Mayer}, E. 1998, Science,
  279, 1332, \dodoi{10.1126/science.279.5355.1332}

\bibitem[{{Herbst} {et~al.}(2020){Herbst}, Gianfranco, \&
  {Ceccarelli}}]{Herbst2020}
{Herbst}, E., Gianfranco, V., \& {Ceccarelli}, C. 2020, ACS Earth and Space
  Chemistry, 4, 488, \dodoi{10.1021/acsearthspacechem.0c00043}

\bibitem[{{Hollis} {et~al.}(2003){Hollis}, {Pedelty}, {Snyder}, {Jewell},
  {Lovas}, {Palmer}, \& {Liu}}]{Hollis03}
{Hollis}, J.~M., {Pedelty}, J.~A., {Snyder}, L.~E., {et~al.} 2003, \apj, 588,
  353

\bibitem[{{Ilyushin} {et~al.}(2013){Ilyushin}, {Endres}, {Lewen}, {Schlemmer},
  \& {Drouin}}]{Ilyushin2013}
{Ilyushin}, V.~V., {Endres}, C.~P., {Lewen}, F., {Schlemmer}, S., \& {Drouin},
  B.~J. 2013, Journal of Molecular Spectroscopy, 290, 31,
  \dodoi{10.1016/j.jms.2013.06.005}

\bibitem[{{Ilyushin} {et~al.}(2021){Ilyushin}, {Margul{\`e}s}, {Tercero},
  {Motiyenko}, {Dorovskaya}, {Alekseev}, {Alonso}, {Kolesnikov{\'a}},
  {Cernicharo}, \& {Guillemin}}]{Ilyushin2021}
{Ilyushin}, V.~V., {Margul{\`e}s}, L., {Tercero}, B., {et~al.} 2021, Journal of
  Molecular Spectroscopy, 379, 111454, \dodoi{10.1016/j.jms.2021.111454}

\bibitem[{{Ioppolo} {et~al.}(2021){Ioppolo}, {Ka{\v{n}}uchov{\'a}}, {James},
  {Dawes}, {Ryabov}, {Dezalay}, {Jones}, {Hoffmann}, {Mason}, \&
  {Strazzulla}}]{Ioppolo2021}
{Ioppolo}, S., {Ka{\v{n}}uchov{\'a}}, Z., {James}, R.~L., {et~al.} 2021, \aap,
  646, A172, \dodoi{10.1051/0004-6361/202039184}

\bibitem[{{Irvine} {et~al.}(1990){Irvine}, {Friberg}, {Kaifu}, {Matthews},
  {Minh}, {Ohishi}, \& {Ishikawa}}]{irvine1990}
{Irvine}, W.~M., {Friberg}, P., {Kaifu}, N., {et~al.} 1990, \aap, 229, L9

\bibitem[{{Jim{\'e}nez-Serra} {et~al.}(2016){Jim{\'e}nez-Serra}, {Vasyunin},
  {Caselli}, {Marcelino}, {Billot}, {Viti}, {Testi}, {Vastel}, {Lefloch}, \&
  {Bachiller}}]{Jimenez-Serra16}
{Jim{\'e}nez-Serra}, I., {Vasyunin}, A.~I., {Caselli}, P., {et~al.} 2016,
  \apjl, 830, L6

\bibitem[{{Jim{\'e}nez-Serra} {et~al.}(2020){Jim{\'e}nez-Serra},
  {Mart{\'\i}n-Pintado}, {Rivilla}, {Rodr{\'\i}guez-Almeida}, {Alonso Alonso},
  {Zeng}, {Cocinero}, {Mart{\'\i}n}, {Requena-Torres}, {Mart{\'\i}n-Domenech},
  \& {Testi}}]{Jimenez-Serra20}
{Jim{\'e}nez-Serra}, I., {Mart{\'\i}n-Pintado}, J., {Rivilla}, V.~M., {et~al.}
  2020, Astrobiology, 20, 1048

\bibitem[{{Jimenez-Serra} {et~al.}(2022){Jimenez-Serra}, {Rodriguez-Almeida},
  {Martin-Pintado}, {Rivilla}, {Melosso}, {Zeng}, {Colzi}, {Kawashima},
  {Hirota}, {Puzzarini}, {Tercero}, {de Vicente}, {Rico-Villas},
  {Requena-Torres}, \& {Martin}}]{jimenez-serra2022}
{Jimenez-Serra}, I., {Rodriguez-Almeida}, L.~F., {Martin-Pintado}, J., {et~al.}
  2022, arXiv e-prints.
\newblock \doarXiv{2204.08267}

\bibitem[{{Jones} {et~al.}(2014){Jones}, {Kaiser}, \& {Strazzulla}}]{Jones2014}
{Jones}, B.~M., {Kaiser}, R.~I., \& {Strazzulla}, G. 2014, \apj, 788, 170,
  \dodoi{10.1088/0004-637X/788/2/170}

\bibitem[{{Jones} {et~al.}(2007){Jones}, {Cunningham}, {Godfrey}, \&
  {Cragg}}]{Jones07}
{Jones}, P.~A., {Cunningham}, M.~R., {Godfrey}, P.~D., \& {Cragg}, D.~M. 2007,
  \mnras, 374, 579

\bibitem[{{J{\o}rgensen} {et~al.}(2018){J{\o}rgensen}, {M{\"u}ller}, {Calcutt},
  {Coutens}, {Drozdovskaya}, {{\"O}berg}, {Persson}, {Taquet}, {van Dishoeck},
  \& {Wampfler}}]{jorgensen2018}
{J{\o}rgensen}, J.~K., {M{\"u}ller}, H.~S.~P., {Calcutt}, H., {et~al.} 2018,
  \aap, 620, A170, \dodoi{10.1051/0004-6361/201731667}

\bibitem[{Kolesniková {et~al.}(2019)Kolesniková, León, Alonso, Mata, \&
  Alonso}]{Kolesnikova19}
Kolesniková, L., León, I., Alonso, E.~R., Mata, S., \& Alonso, J.~L. 2019,
  The Journal of Physical Chemistry Letters, 10, 1325

\bibitem[{Kuan {et~al.}(2003)Kuan, Charnley, Huang, Tseng, \&
  Kisiel}]{Kuan:2003yt}
Kuan, Y.-J., Charnley, S.~B., Huang, H.-C., Tseng, W.-L., \& Kisiel, Z. 2003,
  The Astrophysical Journal, 593, 848

\bibitem[{Lefloch {et~al.}(2017)Lefloch, Ceccarelli, Codella, Favre, Podio,
  Vastel, Viti, \& Bachiller}]{lefloch_l1157-b1_2017}
Lefloch, B., Ceccarelli, C., Codella, C., {et~al.} 2017, Monthly Notices of the
  Royal Astronomical Society: Letters, 469, L73, \dodoi{10.1093/mnrasl/slx050}

\bibitem[{Lester {et~al.}(2001)Lester, Pond, Marshall, Anderson, Harding, \&
  Wagner}]{Lester2001}
Lester, M.~I., Pond, B.~V., Marshall, M.~D., {et~al.} 2001, Faraday Discuss.,
  118, 373, \dodoi{10.1039/B009421H}

\bibitem[{{Liu} {et~al.}(2002){Liu}, {Girart}, {Remijan}, \&
  {Snyder}}]{Liu2002}
{Liu}, S.-Y., {Girart}, J.~M., {Remijan}, A., \& {Snyder}, L.~E. 2002, \apj,
  576, 255, \dodoi{10.1086/341620}

\bibitem[{{Loerting} {et~al.}(2000){Loerting}, {Tautermann}, {Kroemer}, {Kohl},
  {Hallbrucker}, {Mayer}, \& {Liedl}}]{Loerting2000}
{Loerting}, T., {Tautermann}, C., {Kroemer}, R., {et~al.} 2000, Angew. Chem.
  Int. Ed, 39, \dodoi{10.1007/BF00872766}

\bibitem[{{Ma{\c{c}}{\^o}as} {et~al.}(2005){Ma{\c{c}}{\^o}as}, {Khriachtchev},
  {Pettersson}, {Fausto}, \& {R{\"a}s{\"a}nen}}]{Macoas2005}
{Ma{\c{c}}{\^o}as}, E. M.~S., {Khriachtchev}, L., {Pettersson}, M., {Fausto},
  R., \& {R{\"a}s{\"a}nen}, M. 2005, Physical Chemistry Chemical Physics
  (Incorporating Faraday Transactions), 7, 743, \dodoi{10.1039/B416641H}

\bibitem[{{Majumdar} {et~al.}(2018){Majumdar}, {Gratier}, {Wakelam}, {Caux},
  {Willacy}, \& {Ressler}}]{Majumdar2018}
{Majumdar}, L., {Gratier}, P., {Wakelam}, V., {et~al.} 2018, \mnras, 477, 525,
  \dodoi{10.1093/mnras/sty703}

\bibitem[{{Markmeyer} {et~al.}(2019){Markmeyer}, {Lamberts}, {Meisner}, \&
  {K{\"a}stner}}]{Markmeyer2019}
{Markmeyer}, M.~N., {Lamberts}, T., {Meisner}, J., \& {K{\"a}stner}, J. 2019,
  \mnras, 482, 293, \dodoi{10.1093/mnras/sty2620}

\bibitem[{{Mart{\'\i}n} {et~al.}(2019){Mart{\'\i}n}, {Mart{\'\i}n-Pintado},
  {Blanco-S{\'a}nchez}, {Rivilla}, {Rodr{\'\i}guez-Franco}, \&
  {Rico-Villas}}]{martin2019}
{Mart{\'\i}n}, S., {Mart{\'\i}n-Pintado}, J., {Blanco-S{\'a}nchez}, C.,
  {et~al.} 2019, \aap, 631, A159

\bibitem[{Mart\'in {et~al.}(2008)Mart\'in, Requena-Torres, Mart\'in-Pintado, \&
  Mauersberger}]{martin_tracing_2008}
Mart\'in, S., Requena-Torres, M.~A., Mart\'in-Pintado, J., \& Mauersberger, R.
  2008, The Astrophysical Journal, 678, 245

\bibitem[{{McGuire}(2022)}]{McGuire22census}
{McGuire}, B.~A. 2022, \apjs, 259, 30, \dodoi{10.3847/1538-4365/ac2a48}

\bibitem[{{Meg{\'\i}as} {et~al.}(2023){Meg{\'\i}as}, {Jim{\'e}nez-Serra},
  {Mart{\'\i}n-Pintado}, {Vasyunin}, {Spezzano}, {Caselli}, {Cosentino}, \&
  {Viti}}]{Megias2023}
{Meg{\'\i}as}, A., {Jim{\'e}nez-Serra}, I., {Mart{\'\i}n-Pintado}, J., {et~al.}
  2023, \mnras, 519, 1601, \dodoi{10.1093/mnras/stac3449}

\bibitem[{Mehringer {et~al.}(1997)Mehringer, Snyder, Miao, \&
  Lovas}]{Mehringer:1997vk}
Mehringer, D.~M., Snyder, L.~E., Miao, Y., \& Lovas, F.~J. 1997, The
  Astrophysical Journal Letters, 480, L71

\bibitem[{{Molpeceres} \& {Rivilla}(2022)}]{Molpeceres2022}
{Molpeceres}, G., \& {Rivilla}, V.~M. 2022, \aap, 665, A27,
  \dodoi{10.1051/0004-6361/202243892}

\bibitem[{{Molpeceres} {et~al.}(2022){Molpeceres}, {Jim{\'e}nez-Serra}, {Oba},
  {Nguyen}, {Watanabe}, {de la Concepci{\'o}n}, {Mat{\'e}}, {Oliveira}, \&
  {K{\"a}stner}}]{Molpeceres2022a}
{Molpeceres}, G., {Jim{\'e}nez-Serra}, I., {Oba}, Y., {et~al.} 2022, \aap, 663,
  A41, \dodoi{10.1051/0004-6361/202243366}

\bibitem[{{Moore} {et~al.}(2001){Moore}, {Hudson}, \& {Gerakines}}]{Moore2001}
{Moore}, M.~H., {Hudson}, R.~L., \& {Gerakines}, P.~A. 2001, Spectrochimica
  Acta Part A: Molecular Spectroscopy, 57, 843,
  \dodoi{10.1016/S1386-1425(00)00448-0}

\bibitem[{{Mori} {et~al.}(2009{\natexlab{a}}){Mori}, {Suma}, {Sumiyoshi}, \&
  {Endo}}]{Mori2009}
{Mori}, T., {Suma}, K., {Sumiyoshi}, Y., \& {Endo}, Y. 2009{\natexlab{a}},
  \jcp, 130, 204308, \dodoi{10.1063/1.3141405}

\bibitem[{{Mori} {et~al.}(2009{\natexlab{b}}){Mori}, {Suma}, {Sumiyoshi}, \&
  {Endo}}]{Mori09}
---. 2009{\natexlab{b}}, \jcp, 130, 204308

\bibitem[{{Mori} {et~al.}(2011){Mori}, {Suma}, {Sumiyoshi}, \&
  {Endo}}]{Mori2011}
---. 2011, \jcp, 134, 044319, \dodoi{10.1063/1.3532084}

\bibitem[{{M{\"u}ller} {et~al.}(2005){M{\"u}ller}, {Schl{\"o}der}, {Stutzki},
  \& {Winnewisser}}]{Muller2005}
{M{\"u}ller}, H. S.~P., {Schl{\"o}der}, F., {Stutzki}, J., \& {Winnewisser}, G.
  2005, Journal of Molecular Structure, 742, 215,
  \dodoi{10.1016/j.molstruc.2005.01.027}

\bibitem[{{Naraoka} {et~al.}(2023){Naraoka}, {Takano}, {Dworkin}, {Yasuhiro},
  \& {Hamase}}]{Naraoka2023}
{Naraoka}, H., {Takano}, Y., {Dworkin}, J.~P., {Yasuhiro}, O., \& {Hamase}, K.
  2023, Science, 379, \dodoi{10.1126/science.abn9033}

\bibitem[{Nguyen {et~al.}(2012)Nguyen, Xue, Weston, Barker, \&
  Stanton}]{Nguyen2012}
Nguyen, T.~L., Xue, B.~C., Weston, R. E.~J., Barker, J.~R., \& Stanton, J.~F.
  2012, The Journal of Physical Chemistry Letters, 3, 1549,
  \dodoi{10.1021/jz300443a}

\bibitem[{{Noble} {et~al.}(2011){Noble}, {Dulieu}, {Congiu}, \&
  {Fraser}}]{Noble2011}
{Noble}, J.~A., {Dulieu}, F., {Congiu}, E., \& {Fraser}, H.~J. 2011, \apj, 735,
  121, \dodoi{10.1088/0004-637X/735/2/121}

\bibitem[{{Oba} {et~al.}(2010){Oba}, {Watanabe}, {Kouchi}, {Hama}, \&
  {Pirronello}}]{Oba2010}
{Oba}, Y., {Watanabe}, N., {Kouchi}, A., {Hama}, T., \& {Pirronello}, V. 2010,
  \apj, 722, 1598, \dodoi{10.1088/0004-637X/722/2/1598}

\bibitem[{{Pettersson} {et~al.}(2002){Pettersson}, {Ma{\c{c}}{\^o}as},
  {Khriachtchev}, {Lundell}, {Fausto}, \& {R{\"a}s{\"a}nen}}]{Petterson2002}
{Pettersson}, M., {Ma{\c{c}}{\^o}as}, E.~M.~S., {Khriachtchev}, L., {et~al.}
  2002, \jcp, 117, 9095, \dodoi{10.1063/1.1521429}

\bibitem[{Pickett(1991)}]{Pickett:1991cv}
Pickett, H.~M. 1991, Journal of Molecular Spectroscopy, 148, 371

\bibitem[{Pickett {et~al.}(1998)Pickett, Poynter, Cohen, Delitsky, Pearson, \&
  M{\"u}ller}]{1998JQSRT..60..883P}
Pickett, H.~M., Poynter, R.~L., Cohen, E.~A., {et~al.} 1998, Journal of
  Quantitative Spectroscopy and Radiative Transfer, 60, 883

\bibitem[{Pizzarello {et~al.}(2012)Pizzarello, Schrader, Monroe, \&
  Lauretta}]{pizzarello2012}
Pizzarello, S., Schrader, D.~L., Monroe, A.~A., \& Lauretta, D.~S. 2012,
  Proceedings of the National Academy of Sciences, 109, 11949,
  \dodoi{10.1073/pnas.1204865109}

\bibitem[{{Pizzarello} \& {Shock}(2010)}]{Pizzarello2010}
{Pizzarello}, S., \& {Shock}, E. 2010, Cold Spring Harbor perspectives in
  biology, 2.3, \dodoi{10.1101/cshperspect.a002105}

\bibitem[{{Qasim} {et~al.}(2019){Qasim}, {Lamberts}, {He}, {Chuang},
  {Fedoseev}, {Ioppolo}, {Boogert}, \& {Linnartz}}]{Qasim2019}
{Qasim}, D., {Lamberts}, T., {He}, J., {et~al.} 2019, \aap, 626, A118,
  \dodoi{10.1051/0004-6361/201935068}

\bibitem[{Remijan {et~al.}(2003)Remijan, Snyder, Friedel, Liu, \&
  Shah}]{Remijan:2003wf}
Remijan, A., Snyder, L.~E., Friedel, D.~N., Liu, S.-Y., \& Shah, R.~Y. 2003,
  Astrophys. J., 590, 314

\bibitem[{Requena-Torres {et~al.}(2008)Requena-Torres, Mart\'in-Pintado,
  Mart\'in, \& Morris}]{requena-torres_largest_2008}
Requena-Torres, M.~A., Mart\'in-Pintado, J., Mart\'in, S., \& Morris, M.~R.
  2008, The Astrophysical Journal, 672, 352

\bibitem[{Requena-Torres {et~al.}(2006)Requena-Torres, Mart\'in-Pintado,
  Rodr\'iguez-Franco, Mart\'in, Rodr\'iguez-Fern\'andez, \&
  de~Vicente}]{requena-torres_organic_2006}
Requena-Torres, M.~A., Mart\'in-Pintado, J., Rodr\'iguez-Franco, A., {et~al.}
  2006, Astronomy \& Astrophysics, 455, 971, \dodoi{10.1051/0004-6361:20065190}

\bibitem[{Rivilla {et~al.}(2017)Rivilla, Beltr\'an, Mart\'in-Pintado, Fontani,
  Caselli, \& Cesaroni}]{rivilla_chemical_2017}
Rivilla, V.~M., Beltr\'an, M.~T., Mart\'in-Pintado, J., {et~al.} 2017,
  Astronomy and Astrophysics, 599, A26.
\newblock \url{http://adsabs.harvard.edu/abs/2017A%26A...599A..26R}

\bibitem[{{Rivilla} {et~al.}(2019){Rivilla}, {Mart{\'\i}n-Pintado},
  {Jim{\'e}nez-Serra}, {Zeng}, {Mart{\'\i}n}, {Armijos-Abenda{\~n}o},
  {Requena-Torres}, {Aladro}, \& {Riquelme}}]{rivilla2019b}
{Rivilla}, V.~M., {Mart{\'\i}n-Pintado}, J., {Jim{\'e}nez-Serra}, I., {et~al.}
  2019, \mnras, 483, L114

\bibitem[{{Rivilla} {et~al.}(2020){Rivilla}, {Mart{\'\i}n-Pintado},
  {Jim{\'e}nez-Serra}, {Mart{\'\i}n}, {Rodr{\'\i}guez-Almeida},
  {Requena-Torres}, {Rico-Villas}, {Zeng}, \& {Briones}}]{rivilla2020b}
---. 2020, \apjl, 899, L28

\bibitem[{{Rivilla} {et~al.}(2021{\natexlab{a}}){Rivilla}, {Jim{\'e}nez-Serra},
  {Mart{\'\i}n-Pintado}, {Briones}, {Rodr{\'\i}guez-Almeida}, {Rico-Villas},
  {Tercero}, {Zeng}, {Colzi}, {de Vicente}, {Mart{\'\i}n}, \&
  {Requena-Torres}}]{rivilla2021a}
{Rivilla}, V.~M., {Jim{\'e}nez-Serra}, I., {Mart{\'\i}n-Pintado}, J., {et~al.}
  2021{\natexlab{a}}, Proceedings of the National Academy of Science, 118,
  2101314118

\bibitem[{{Rivilla} {et~al.}(2021{\natexlab{b}}){Rivilla}, {Jim{\'e}nez-Serra},
  {Garc{\'\i}a de la Concepci{\'o}n}, {Mart{\'\i}n-Pintado}, {Colzi},
  {Rodr{\'\i}guez-Almeida}, {Tercero}, {Rico-Villas}, {Zeng}, {Mart{\'\i}n},
  {Requena-Torres}, \& {de Vicente}}]{rivilla2021b}
{Rivilla}, V.~M., {Jim{\'e}nez-Serra}, I., {Garc{\'\i}a de la Concepci{\'o}n},
  J., {et~al.} 2021{\natexlab{b}}, \mnras, 506, L79

\bibitem[{{Rivilla} {et~al.}(2022{\natexlab{a}}){Rivilla}, {Colzi},
  {Jim{\'e}nez-Serra}, {Mart{\'\i}n-Pintado}, {Meg{\'\i}as}, {Melosso},
  {Bizzocchi}, {L{\'o}pez-Gallifa}, {Mart{\'\i}nez-Henares}, {Massalkhi},
  {Tercero}, {de Vicente}, {Guillemin}, {Garc{\'\i}a de la Concepci{\'o}n},
  {Rico-Villas}, {Zeng}, {Mart{\'\i}n}, {Requena-Torres}, {Tonolo},
  {Alessandrini}, {Dore}, {Barone}, \& {Puzzarini}}]{rivilla2022a}
{Rivilla}, V.~M., {Colzi}, L., {Jim{\'e}nez-Serra}, I., {et~al.}
  2022{\natexlab{a}}, \apjl, 929, L11

\bibitem[{{Rivilla} {et~al.}(2022{\natexlab{b}}){Rivilla}, {Garc{\'\i}a de la
  Concepci{\'o}n}, {Jim{\'e}nez-Serra}, {Mart{\'\i}n-Pintado}, {Colzi},
  {Tercero}, {Meg{\'\i}as}, {L{\'o}pez-Gallifa}, {Mart{\'\i}nez-Henares},
  {Massalkhi}, {Mart{\'\i}n}, {Zeng}, {De Vicente}, {Rico-Villas},
  {Requena-Torres}, \& {Cosentino}}]{rivilla2022b}
{Rivilla}, V.~M., {Garc{\'\i}a de la Concepci{\'o}n}, J., {Jim{\'e}nez-Serra},
  I., {et~al.} 2022{\natexlab{b}}, arXiv e-prints

\bibitem[{{Rodr{\'\i}guez-Almeida}
  {et~al.}(2021{\natexlab{a}}){Rodr{\'\i}guez-Almeida}, {Jim{\'e}nez-Serra},
  {Rivilla}, {Mart{\'\i}n-Pintado}, {Zeng}, {Tercero}, {de Vicente}, {Colzi},
  {Rico-Villas}, {Mart{\'\i}n}, \& {Requena-Torres}}]{rodriguez-almeida2021a}
{Rodr{\'\i}guez-Almeida}, L.~F., {Jim{\'e}nez-Serra}, I., {Rivilla}, V.~M.,
  {et~al.} 2021{\natexlab{a}}, \apjl, 912, L11

\bibitem[{{Rodr{\'\i}guez-Almeida}
  {et~al.}(2021{\natexlab{b}}){Rodr{\'\i}guez-Almeida}, {Rivilla},
  {Jim{\'e}nez-Serra}, {Melosso}, {Colzi}, {Zeng}, {Tercero}, {de Vicente},
  {Mart{\'\i}n}, {Requena-Torres}, {Rico-Villas}, \&
  {Mart{\'\i}n-Pintado}}]{rodriguez-almeida2021b}
{Rodr{\'\i}guez-Almeida}, L.~F., {Rivilla}, V.~M., {Jim{\'e}nez-Serra}, I.,
  {et~al.} 2021{\natexlab{b}}, \aap, 654, L1

\bibitem[{{Ruaud} {et~al.}(2015){Ruaud}, {Loison}, {Hickson}, {Gratier},
  {Hersant}, \& {Wakelam}}]{Ruad2015}
{Ruaud}, M., {Loison}, J.~C., {Hickson}, K.~M., {et~al.} 2015, \mnras, 447,
  4004, \dodoi{10.1093/mnras/stu2709}

\bibitem[{Rubin {et~al.}(2019)Rubin, Bekaert, Broadley, Drozdovskaya, \&
  Wampfler}]{rubin2019}
Rubin, M., Bekaert, D.~V., Broadley, M.~W., Drozdovskaya, M.~N., \& Wampfler,
  S.~F. 2019, ACS Earth and Space Chemistry, 3, 1792,
  \dodoi{10.1021/acsearthspacechem.9b00096}

\bibitem[{{Sanz-Novo} {et~al.}(2021){Sanz-Novo}, {Le{\'o}n}, {Alonso},
  {Kolesnikov{\'a}}, \& {Alonso}}]{Sanz-Novo21}
{Sanz-Novo}, M., {Le{\'o}n}, I., {Alonso}, E.~R., {Kolesnikov{\'a}}, L., \&
  {Alonso}, J.~L. 2021, \apj, 915, 76

\bibitem[{{Sephton}(2002)}]{Sephton2002}
{Sephton}, M.~A. 2002, Natural Product Reports, 19, 292,
  \dodoi{10.1039/b103775g}

\bibitem[{{Snyder} {et~al.}(2005){Snyder}, {Lovas}, {Hollis}, {Friedel},
  {Jewell}, {Remijan}, {Ilyushin}, {Alekseev}, \& {Dyubko}}]{Snyder05}
{Snyder}, L.~E., {Lovas}, F.~J., {Hollis}, J.~M., {et~al.} 2005, \apj, 619, 914

\bibitem[{{Strazzulla} {et~al.}(1996){Strazzulla}, {Brucato}, {Cimino}, \&
  {Palumbo}}]{Strazzulla1996}
{Strazzulla}, G., {Brucato}, J.~R., {Cimino}, G., \& {Palumbo}, M.~E. 1996,
  \planss, 44, 1447, \dodoi{10.1016/S0032-0633(96)00079-7}

\bibitem[{Tachikawa \& Kawabata(2016)}]{Tachikawa2016}
Tachikawa, H., \& Kawabata, H. 2016, The Journal of Physical Chemistry A, 120,
  6596, \dodoi{10.1021/acs.jpca.6b05563}

\bibitem[{{Taquet} {et~al.}(2017){Taquet}, {Wirstr{\"o}m}, {Charnley}, {Faure},
  {L{\'o}pez-Sepulcre}, \& {Persson}}]{taquet2017}
{Taquet}, V., {Wirstr{\"o}m}, E.~S., {Charnley}, S.~B., {et~al.} 2017, \aap,
  607, A20, \dodoi{10.1051/0004-6361/201630023}

\bibitem[{{Tercero} {et~al.}(2018){Tercero}, {Cuadrado}, {L{\'o}pez},
  {Brouillet}, {Despois}, \& {Cernicharo}}]{Tercero2018}
{Tercero}, B., {Cuadrado}, S., {L{\'o}pez}, A., {et~al.} 2018, \aap, 620, L6,
  \dodoi{10.1051/0004-6361/201834417}

\bibitem[{{Tercero} {et~al.}(2021){Tercero}, {L{\'o}pez-P{\'e}rez}, {Gallego},
  {Beltr{\'a}n}, {Garc{\'\i}a}, {Patino-Esteban}, {L{\'o}pez-Fern{\'a}ndez},
  {G{\'o}mez-Molina}, {Diez}, {Garc{\'\i}a-Carre{\~n}o}, {Malo}, {Amils},
  {Serna}, {Albo}, {Hern{\'a}ndez}, {Vaquero}, {Gonz{\'a}lez-Garc{\'\i}a},
  {Barbas}, {L{\'o}pez-Fern{\'a}ndez}, {Bujarrabal}, {G{\'o}mez-Garrido},
  {Pardo}, {Santander-Garc{\'\i}a}, {Tercero}, {Cernicharo}, \& {de
  Vicente}}]{tercero2021}
{Tercero}, F., {L{\'o}pez-P{\'e}rez}, J.~A., {Gallego}, J.~D., {et~al.} 2021,
  \aap, 645, A37, \dodoi{10.1051/0004-6361/202038701}

\bibitem[{{Thaddeus} {et~al.}(1981){Thaddeus}, {Guelin}, \&
  {Linke}}]{Thaddeus1981}
{Thaddeus}, P., {Guelin}, M., \& {Linke}, R.~A. 1981, \apjl, 246, L41,
  \dodoi{10.1086/183549}

\bibitem[{{Tsuge} \& {Khriachtchev}(2015)}]{Tsuge2015}
{Tsuge}, M., \& {Khriachtchev}, L. 2015, Journal of Physical Chemistry A, 119,
  2628, \dodoi{10.1021/jp509692b}

\bibitem[{Wagner {et~al.}(2016)Wagner, Reisenauer, Hirvonen, Wu, Tyberg, Allen,
  \& Schreiner}]{Wagner2016}
Wagner, J.~P., Reisenauer, H.~P., Hirvonen, V., {et~al.} 2016, Chem. Commun.,
  52, 7858, \dodoi{10.1039/C6CC01756H}

\bibitem[{{Wang} {et~al.}(2016){Wang}, {Zeuschner}, {Eremets}, {Troyan}, \&
  {Willams}}]{Wang2016}
{Wang}, H., {Zeuschner}, J., {Eremets}, M., {Troyan}, I., \& {Willams}, J.
  2016, Scientific Reports, 6, 19902, \dodoi{10.1038/srep19902}

\bibitem[{{Winnewisser} {et~al.}(2002){Winnewisser}, {Winnewisser}, {Stein},
  {Birk}, {Wagner}, {Winnewisser}, {Yamada}, {Belov}, \&
  {Baskakov}}]{Winnewisser2002}
{Winnewisser}, M., {Winnewisser}, B.~P., {Stein}, M., {et~al.} 2002, Journal of
  Molecular Spectroscopy, 216, 259, \dodoi{10.1006/jmsp.2002.8590}

\bibitem[{{Zeng} {et~al.}(2018){Zeng}, {Jim{\'e}nez-Serra}, {Rivilla},
  {Mart{\'{\i}}n}, {Mart{\'{\i}}n-Pintado}, {Requena-Torres},
  {Armijos-Abenda{\~n}o}, {Riquelme}, \& {Aladro}}]{zeng2018}
{Zeng}, S., {Jim{\'e}nez-Serra}, I., {Rivilla}, V.~M., {et~al.} 2018, Monthly
  Notices of the Royal Astronomical Society, 478, 2962

\bibitem[{{Zeng} {et~al.}(2020){Zeng}, {Zhang}, {Jim{\'e}nez-Serra}, {Tercero},
  {Lu}, {Mart{\'\i}n-Pintado}, {de Vicente}, {Rivilla}, \& {Li}}]{zeng2020}
{Zeng}, S., {Zhang}, Q., {Jim{\'e}nez-Serra}, I., {et~al.} 2020, \mnras, 497,
  4896.
\newblock \doarXiv{2007.14362}

\bibitem[{{Zeng} {et~al.}(2021){Zeng}, {Jim{\'e}nez-Serra}, {Rivilla},
  {Mart{\'\i}n-Pintado}, {Rodr{\'\i}guez-Almeida}, {Tercero}, {de Vicente},
  {Rico-Villas}, {Colzi}, {Mart{\'\i}n}, \& {Requena-Torres}}]{zeng2021}
{Zeng}, S., {Jim{\'e}nez-Serra}, I., {Rivilla}, V.~M., {et~al.} 2021, \apjl,
  920, L27

\bibitem[{{Zeng} {et~al.}(2023){Zeng}, {Rivilla}, {Jim{\'e}nez-Serra}, {Colzi},
  {Mart{\'\i}n-Pintado}, {Tercero}, {de Vicente}, {Mart{\'\i}n}, \&
  {Requena-Torres}}]{zeng2023}
{Zeng}, S., {Rivilla}, V.~M., {Jim{\'e}nez-Serra}, I., {et~al.} 2023, \mnras,
  523, 1448, \dodoi{10.1093/mnras/stad1478}

\bibitem[{{Zheng} \& {Kaiser}(2007)}]{Zheng2007}
{Zheng}, W., \& {Kaiser}, R.~I. 2007, Chemical Physics Letters, 450, 55,
  \dodoi{10.1016/j.cplett.2007.10.094}

\bibitem[{{Zhu} {et~al.}(2018){Zhu}, {Turner}, {Abplanalp}, \&
  {Kaiser}}]{Zhu2018}
{Zhu}, C., {Turner}, A.~M., {Abplanalp}, M.~J., \& {Kaiser}, R.~I. 2018, \apjs,
  234, 15, \dodoi{10.3847/1538-4365/aa9f28}

\bibitem[{Zuckerman {et~al.}(1971)Zuckerman, Ball, \&
  Gottlieb}]{Zuckerman:1971de}
Zuckerman, B., Ball, J.~A., \& Gottlieb, C.~A. 1971, The Astrophysical Journal,
  163, L41

\end{thebibliography}
\bibliographystyle{aasjournal}

\newpage
\appendix
\twocolumngrid

\section{Spectroscopic details and complementary tables}
\label{Comtable}

\restartappendixnumbering

In Table \ref{t:mwdata} we list the experimental constants for the ground state of \textit{cis-cis} and \textit{cis-trans} conformers of carbonic acid. We report in the third column the results of the new global fit, including lines from the astronomical data correctly weighted, which have been measured for the first time in space.

Regarding the \textit{cis-cis} conformer of HOCOOH (C$_{2v}$ symmetry), as reported in \citealt{Mori2011}, the rotational energy levels must be evaluated carefully since this conformer has one pair of equivalent protons (obeying Fermi–Dirac statistics). Therefore, the existence of nuclear statistics are key in the determination of the relative intensities of rotational transitions; the levels will be divided in those where the rotational functions are symmetric with respect to a rotation around the \textit{b}-symmetry axis (\textit{K$_a$}, \textit{K$_c$} equals “even, even” or “odd, odd”, para levels) and those with symmetric rotational functions (\textit{K$_a$}, \textit{K$_c$} equals “even, odd” or “odd, even”, ortho levels). Thus, the proper nuclear statistical weight is 1:3 and, at low temperatures, both ortho and para levels should be analyzed separately. 

\begin{table}
\begin{center}
\setlength{\tabcolsep}{1.4pt}
\caption{Experimental spectroscopic parameters for the ground state of \textit{cis-cis} (cc) and \textit{cis-trans} (ct) HOCOOH ($A$-Reduction, I$^{r}$-Representation)}
\label{t:mwdata}
\vspace*{0.0ex}
\begin{tabular}{lllll}
\hline\hline
\multicolumn{1}{c}{Parameter} & \multicolumn{1}{c}{\textit{cis-cis} (cc) $^{(f)}$} & \multicolumn{1}{c}{\textit{cis-trans} (ct) $^{(f)}$} & \multicolumn{1}{c}{Global Fit (ct) $^{(h)}$} \\ 
\hline
$A^{(a)}$\small & 11997.0646 (22) & 11778.6808 (34) & 11778.6801 (31) \\
$B$ \small  & 11308.3818 (14) & 11423.1345 (32) & 11423.1340 (30)  \\ 
$C$ \small  & 5813.82769 (71) &  5792.0741 (22) & 5792.0740 (21)  \\ 
$\Delta_J$ \small & 6.371 (72) &  5.74 (18) & 5.64 (18) \\ 
$\Delta_{JK}$ \small  & –2.13 (95) & -1.16 (79) & -1.14 (72)  \\ 
$\Delta_K$ \small  & 6.42 (79) & 8.14 (74) & 8.18 (68)  \\ 
$\delta_J$ \small & 2.618 $^{(g)}$ & 2.618 (83) & 2.616 (77)  \\ 
$\delta_K$ \small  & 6.28 & 6.28 (31) & 6.30 (28)  \\ 
$\Phi_J$ \small  & - &- & 5.3 (15)  \\ 
$N$ $^{(b)}$ \small  & 7 & 25 & 41 \\ 
$\sigma^{(c)}$\small & 1.0 & 8.6 & 25.8 \\ 
$\sigma_{w}^{(d)}$\small & 0.98 & 0.86 & 0.81 \\ 
$\Delta$E$^{(e)}$\small & 0.0 & 1.71 & 1.71  \\ 
\hline 
\end{tabular}
\end{center}
\vspace*{-0.5ex}
\tablecomments{Standard error in parentheses are displayed in units of the last digit. $^{(a)}$$A$, $B$, and $C$ represent the rotational constants (in MHz), while $\Delta_J$, $\Delta_{JK}$, $\Delta_J$, $\delta_J$ and $\delta_K$ are the quartic centrifugal distortion constants (in kHz), and $\Phi_J$ is a sextic centrifugal distortion constant (in Hz). $^{(b)}$ $N$ is the number of measured transitions. $^{(c)}$ $\sigma$ is the root mean square (RMS) deviation of the fit (kHz). $^{(d)}$$\sigma_{w}$ is the unitless (weighted) deviation of the fit. Their values are close to the 1$\sigma$ standard uncertainties since the unitless (weighted) deviation of the fit is very close to 1.0.$^{(e)}$ $\Delta$$E$ is the energy calculated at the CCSD(T)/cc-pVQZ level (in kcal mol $^{-1}$, \citealt{Mori09}). $^{(f)}$  Spectroscopic parameters derived from the frequency measurements reported in \citet{Mori09,Mori2011}. $^{(g)}$ Values fixed to that of \textit{cis-trans} carbonic acid. $^{(h)}$ Global fit of \textit{cis-trans} carbonic acid where newly measured astronomical lines are also included.}
\end{table}

In Table \ref{t:pfun} we provide the rotational ($Q$$_r$) partition function of the ground state of conformers \textit{cis-cis} and \textit{cis-trans}, which are mandatory to obtain reliable line intensities. We used SPCAT \citep{Pickett:1991cv} to estimate the values of $Q$$_r$ from first principles at the conventional temperatures as implemented in the JPL database \citep{1998JQSRT..60..883P}, and an additional temperature of 5 K, using the rotational spectroscopic parameters reported in Table \ref{t:mwdata} and Eq. 3.67 of \cite{Gordy:1984uy}. An example table (Table \ref{t:cat}) for the generated ctHOCOOH.cat file is also provided.

\begin{table}
\begin{center}
\caption{Rotational partition function of \textit{cis-cis} (cc) and \textit{cis-trans} (ct) carbonic acid.}
\label{t:pfun}
\vspace*{0.0ex}
\begin{tabular}{ccccc}
\hline\hline
\multicolumn{1}{c}{$T$}{\small (K)} & \multicolumn{1}{c}{$Q$$_r$(cc)} & \multicolumn{1}{c}{log$Q$$_r$(cc)} & \multicolumn{1}{c}{$Q$$_r$(ct)} & \multicolumn{1}{c}{log$Q$$_r$(ct)}\\ 
\hline
5.000 & 68.3830 & 1.8349 & 68.7904 & 1.8375 \\
9.375 & 174.039 & 2.2406 & 175.082 & 2.2432\\
18.75 & 489.828 & 2.6900 & 492.771 & 2.6926 \\ 
37.50 & 1382.12 & 3.1405 & 1390.43 & 3.1432 \\ 
75.00 & 3905.06 & 3.5916 & 3928.54 &  3.5942 \\
150.0 & 11042.1 & 4.0431 & 11108.4 &  4.0457 \\ 
225.0 & 20282.4 & 4.3071 & 20409.2 &  4.3098 \\ 
300.0 & 31188.0 & 4.4940 & 31430.9 &  4.4974 \\ 
\hline 
\end{tabular}
\vspace*{-0.5ex}
\tablecomments{We used $J$ = 200 as the maximum $J$ value.}
\end{center}
\end{table}

\begin{table*}
\begin{center}
\caption{Example table of the .cat file provided for \textit{cis-trans} HOCOOH (ctHOCOOH.cat) in the JPL SPFIT/SPCAT format.}
\label{t:cat}
\vspace*{0.0ex}
\begin{tabular}{cccccccccc}
\hline\hline
Frequency & Error $^{(a)}$ & log \textit{Int} $^{(b)}$& DR $^{(c)}$ & E$_{LO}$ $^{(d)}$ & \textit{g}$_u$ $^{(e)}$ & TAG $^{(f)}$ & Q$_N$FMT $^{(g)}$&  Q$_N$' $^{(h)}$&  Q$_N$" $^{(i)}$ \\ 
(MHz)  &  (MHz)  &  (nm$^2$ MHz) & & (cm$^{-1}$) &   & & \\
\hline
   40567.0170  & 0.0046 & -6.0610 & 3  &  1.5462 &  7 & 62994 & 303 & 3 0 3   &    2 0 2    \\    
   40567.4840  & 0.0047 & -5.1428 & 3  &  1.5462 & 7  & 62994 & 303 & 3 1 3   &    2 0 2    \\   
   75310.2432  & 0.0585 & -4.2732 & 3  &  6.7644 &13  & 62994 & 303 & 6 0 6   &    5 1 5    \\    
   75310.2432  & 0.0585 & -5.1953 & 3  &  6.7644 &13  & 62994 & 303 & 6 1 6   &    5 1 5    \\    
   75310.2435  & 0.0585 & -5.1953 & 3  &  6.7644 &13  & 62994 & 303 & 6 0 6   &    5 0 5    \\    
   75310.2435  & 0.0585 & -4.2732 & 3  &  6.7644 &13  & 62994 & 303 & 6 1 6   &    5 0 5    \\    
   75338.1091  & 0.0372 & -4.4608 & 3  &  5.9940 &11  & 62994 & 303 & 5 1 4   &    4 2 3    \\    
   75338.2086  & 0.0372 & -5.3723 & 3  &  5.9940 &11  & 62994 & 303 & 5 2 4   &    4 2 3    \\    
   75341.3737  & 0.0372 & -5.3722 & 3  &  5.9939 &11  & 62994 & 303 & 5 1 4   &    4 1 3    \\    
   75341.4732  & 0.0372 & -4.4607 & 3  &  5.9939 &11  & 62994 & 303 & 5 2 4   &    4 1 3    \\ 
\hline 
\end{tabular}
\vspace*{-0.5ex}
\tablecomments{$^{(a)}$Estimated or experimental error. $^{(b)}$Base 10 logarithm of the integrated intensity at 300 K.$^{(c)}$Degrees of freedom in the rotational partition function. $^{(d)}$Lower state energy relative to the ground state. $^{(e)}$Upper state degeneracy. $^{(f)}$Molecular tag or species identifier. $^{(g)}$Identifies the format of the quantum numbers. $^{(h)}$Quantum numbers for the upper state. $^{(i)}$Quantum numbers for the lower state.}
\end{center}
\end{table*}

\section{Analysis of acetic acid toward the G+0.693-0.027 molecular cloud}
\label{AnalysisCH3COOH}

\begin{center}
\begin{figure*}[ht]
     \centerline{\resizebox{1.05
     \hsize}{!}{\includegraphics[angle=0]{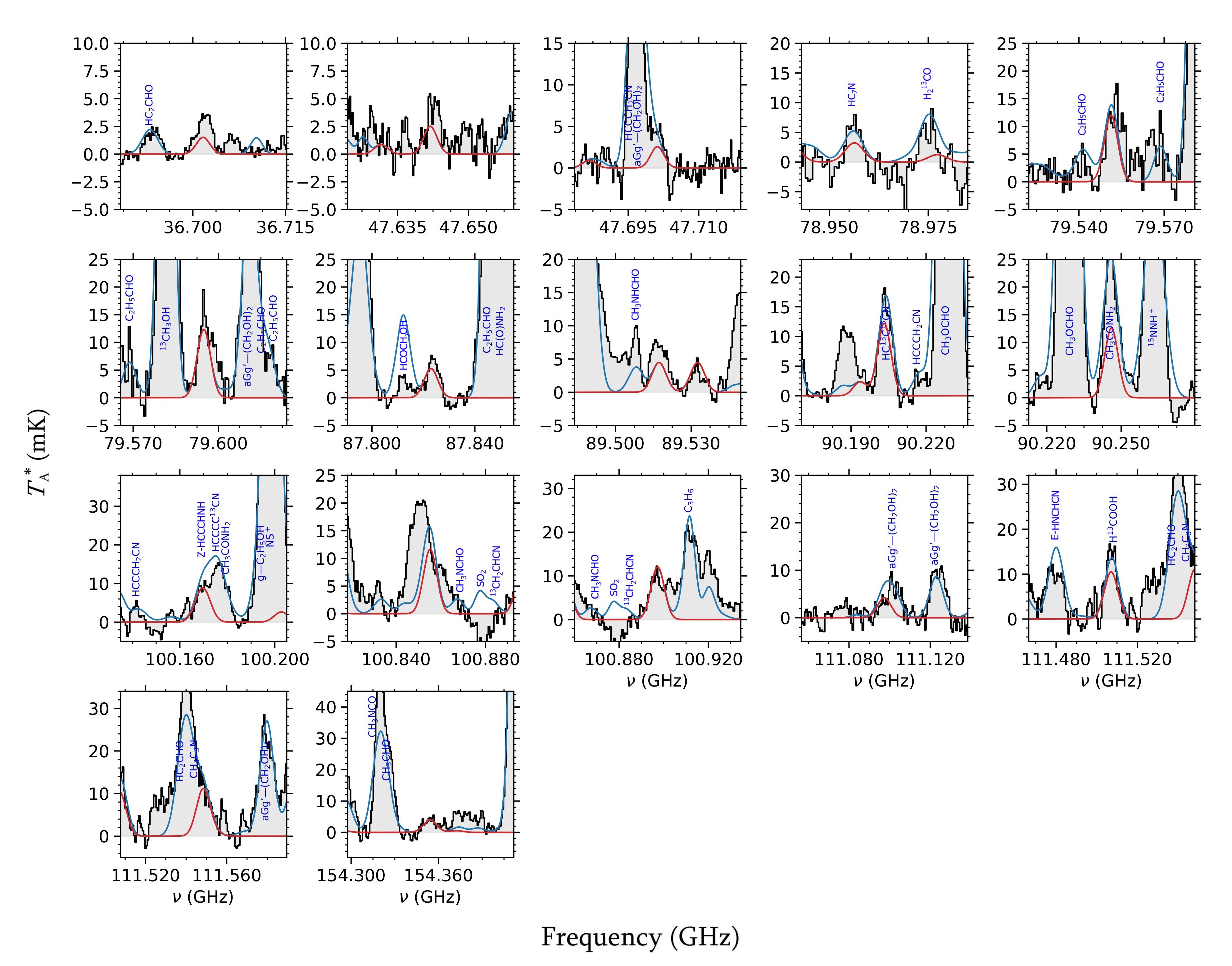}}}
     \caption{Selected transitions of CH$_{3}$COOH identified toward the G+0.693–0.027 molecular cloud. The best LTE fit computed with \textsc{Madcuba-Slim} is depicted in a red solid line and the predicted molecular emission from all of the molecules identified in our spectral survey is shown in blue (observed spectra plotted as gray histograms).}
\label{f:LTECH3COOH}
\end{figure*}
\end{center}

To carry out the corresponding LTE analysis of  CH$_{3}$COOH, as well as to constrain the excitation conditions, we picked transitions that: i) are not significantly blended with other molecular species and ii) cover a relatively wide range of rotational energy levels (see Table \ref{t:CH3COOH}). We employed the set of spectroscopic parameters of the ground vibrational state ($v$$_t$ = 0) of CH$_{3}$COOH reported in \citet{Ilyushin2013} (entry 60523 of the CDMS catalog; \citealt{Muller2005}). We again used MADCUBA, leaving as free parameters: the molecular column density ($N$), the excitation temperature ($T_{\rm ex}$) and the radial velocity ($v$$_{\rm LSR}$). The value of the linewidth (FWHM) was fixed to 21 kms s$^{-1}$ in the fit. In Table \ref{tab:comparisonacids} we present the derived physical parameters of CH$_{3}$COOH along with the physical parameters of formic acid and carbonic acid, while the results of the LTE fits of the transitions are depicted in Figure \ref{f:LTECH3COOH}.

\begin{table*}
\centering
\tabcolsep 3pt
\caption{Spectroscopic information of the selected unblended or slightly blended transitions of CH$_{3}$COOH detected toward G+0.693$-$0.027 (shown in Fig. \ref{f:LTECH3COOH})}
\begin{tabular}{ccccccccc}
\hline\hline
Frequency & Transition $^{(a)}$ & log \textit{I}& \textit{g}$_u$ & E$_{LO}$ & E$_{up}$ & Comments  \\ 
(GHz) & &  (nm$^2$ MHz) & & (cm$^{-1}$) &  (K ) & & \\
\hline
36.7017105 & 3$_{0,3}$-$2_{1,2}$ E  & -6.6938 & 7 & 1.8 & 4.3 & Unblended\\      
47.6419985 & 4$_{1,4}$-$3_{0,3}$ E  & -6.3107 & 9 & 3.0 & 6.5 & Unblended \\   
47.7012763 & 4$_{1,4}-$3$_{0,3}$ A  & -6.3130 & 9 & 2.6 & 6.0 & Blended: $aGg'$-(CH$_2$OH)$_2$\\   
78.9564732 & 6$_{1,5}-$5$_{2,4}$ A  & -5.7912 & 13 & 7.7 & 14.7 & Blended: HC$_7$N\\  
79.5513516 & 7$_{0,7}-$6$_{1,6}$ E  & -5.5998 & 15 & 8.8 & 16.4 & Unblended\\ 
79.5513690 & 7$_{1,7}-$6$_{1,6}$ E  & -6.0844 & 15 & 8.8 & 16.4 & Unblended\\ 
79.5515061 & 7$_{0,7}-$6$_{0,6}$ E  & -6.0844 & 15 & 8.8 & 16.4 & Unblended\\  
79.5515235 & 7$_{1,7}-$6$_{0,6}$ E  & -5.5998 & 15 & 8.8 & 16.4 & Unblended\\ 
79.5948832 & 7$_{0,7}-$6$_{1,6}$ A  & -5.6048 & 15 & 8.4 & 15.9 & Unblended\\  
79.5949136 & 7$_{1,7}-$6$_{1,6}$ A  & -6.0684 & 15 & 8.4 & 15.9 & Unblended\\  
79.5951321 & 7$_{0,7}-$6$_{0,6}$ A  & -6.0684 & 15 & 8.4 & 15.9 & Unblended\\  
79.5951625 & 7$_{1,7}-$6$_{0,6}$ A  & -5.6048 & 15 & 8.4 & 15.9 & Unblended\\  
87.8236760 & 6$_{2,4}-$5$_{3,3}$ A  & -5.8593 & 13 & 8.8 & 16.8 & Unblended \\  
89.5169537 & 7$_{1,6}-$6$_{2,5}$ E  & -5.5899 & 15 & 10.6 & 19.5& Unblended \\  
89.5311839 & 7$_{1,6}-$6$_{1,5}$ E  & -6.0356 & 15 & 10.6 & 19.5 & Unblended \\    
90.2034364 & 8$_{0,8}-$7$_{1,7}$ E  & -5.4336 & 17 & 11.5 & 20.7 & Blended: HCC$^{13}$C$^{13}$N\\ 
90.2034383 & 8$_{1,8}-$7$_{1,7}$ E  & -5.9223 & 17 & 11.5 & 20.7 & Blended: HCC$^{13}$C$^{13}$N\\ 
90.2034538 & 8$_{0,8}-$7$_{0,7}$ E  & -5.9223 & 17 & 11.5 & 20.7 & Blended: HCC$^{13}$C$^{13}$N\\ 
90.2034556 & 8$_{1,8}-$7$_{0,7}$ E  & -5.4336 & 17 & 11.5 & 20.7 & Blended: HCC$^{13}$C$^{13}$N\\ 
90.2462358 & 8$_{0,8}-$7$_{1,7}$ A  & -5.4378 & 17 & 11.1 & 20.2 & Blended: CH$_3$CONH$_2$\\ 
90.2462394 & 8$_{1,8}-$7$_{1,7}$ A  & -5.9027 & 17 & 11.1 & 20.2 & Blended: CH$_3$CONH$_2$\\  
90.2462662 & 8$_{0,8}-$7$_{0,7}$ A  & -5.9027 & 17 & 11.1 & 20.2 & Blended: CH$_3$CONH$_2$\\  
90.2462697 & 8$_{1,8}-$7$_{0,7}$ A  & -5.4378 & 17 & 11.1 & 20.2 & Blended: CH$_3$CONH$_2$\\  
100.1687044 & 8$_{1,7}-$7$_{2,6}$ E  & -5.4228 & 17 & 13.6 & 24.2 & Blended: CH$_3$CONH$_2$, HC$_4$$^{13}$CN and Z-HCCCHNH\\  
100.1689638 & 8$_{2,7}-$7$_{2,6}$ E  & -5.8783 & 17 & 13.6 & 24.2 & Blended: CH$_3$CONH$_2$, HC$_4$$^{13}$CN and Z-HCCCHNH\\  
100.1706980 & 8$_{1,7}-$7$_{1,6}$ E  & -5.8783 & 17 & 13.6 & 24.2 & Blended: CH$_3$CONH$_2$, HC$_4$$^{13}$CN \\  
100.1709574 & 8$_{2,7}-$7$_{1,6}$ E  & -5.4228 & 17 & 13.6 & 24.2 & Blended: CH$_3$CONH$_2$, HC$_4$$^{13}$CN \\  
100.8554272 & 9$_{0,9}-$8$_{1,8}$ E  & -5.2883 & 19 & 14.5 & 25.5 & Blended: CH$_3$NCHO and U-line\\  
100.8554274 & 9$_{1,9}-$8$_{1,8}$ E  & -5.7817 & 19 & 14.5 & 25.5 & Blended: CH$_3$NCHO and U-line\\  
100.8554291 & 9$_{0,9}-$8$_{0,8}$ E  & -5.7817 & 19 & 14.5 & 25.5 & Blended: CH$_3$NCHO and U-line\\  
100.8554293 & 9$_{1,9}-$8$_{0,8}$ E  & -5.2883 & 19 & 14.5 & 25.5 & Blended: CH$_3$NCHO and U-line\\  
100.8974541 & 9$_{0,9}-$8$_{1,8}$ A  & -5.2943 & 19 & 14.1 & 25.0 & Unblended \\ 
100.8974545 & 9$_{1,9}-$8$_{1,8}$ A  & -5.7585 & 19 & 14.1 & 25.0 & Unblended\\ 
100.8974577 & 9$_{0,9}-$8$_{0,8}$ A  & -5.7585 & 19 & 14.1 & 25.0 & Unblended\\ 
100.8974581 & 9$_{1,9}-$8$_{0,8}$ A  & -5.2943 & 19 & 14.1 & 25.0 & Unblended\\    
111.0975729 & 5$_{5,0}-$4$_{1,1}$ A  & -5.5160 & 11 &  7.2 & 15.5 & Blended: $aGg'$-(CH$_2$OH)$_2$\\ 
111.5072803 &10$_{0,10}-$9$_{1,9}$ E  & -5.1575 & 21 & 17.8 & 30.8  & Unblended \\  
111.5072803 &10$_{1,10}-$9$_{1,9}$ E  &  -5.6580 & 21 & 17.8 & 30.8 & Unblended \\  
111.5072805 &10$_{1,10}-$9$_{0,9}$ E  &  -5.1575 & 21 & 17.8 & 30.8 & Unblended \\  
111.5072805 &10$_{0,10}-$9$_{0,9}$ E  &  -5.6580 & 21 & 17.8 & 30.8 & Unblended \\  
111.5485353 &10$_{0,10}-$9$_{1,9}$ A  &  -5.1649 & 21 & 17.5 & 30.3 & Blended: CH$_3$C$_3$N \\  
111.5485353 &10$_{1,10}-$9$_{1,9}$ A  &  -5.6307 & 21 & 17.5 & 30.3 & Blended: CH$_3$C$_3$N\\  
111.5485357 &10$_{1,10}-$9$_{0,9}$ A  &  -5.1649 & 21 & 17.5 & 30.3 & Blended: CH$_3$C$_3$N\\  
111.5485357 &10$_{0,10}-$9$_{0,9}$ A  &  -5.6307 & 21 & 17.5 & 30.3 & Blended: CH$_3$C$_3$N\\ 
154.3545087 & 7$_{7,1}-$6$_{6,1}$ E  &  -5.0690 & 15 & 15.4 & 29.3 & Unblended\\ 
\hline 
\end{tabular}
\label{t:CH3COOH}
\vspace*{-2.5ex}
\tablecomments{$^{(a)}$ The rotational energy levels are labelled using the conventional notation for asymmetric tops: $J_{K_{a},K_{c}}$, where $J$ denotes the angular momentum quantum number, and the $K_{a}$ and $K_{c}$ labels are projections of $J$ along the $a$ and $c$ principal axes. The $A$ and $E$ labels refer to the $A$ and $E$ substates, respectively, originated from the methyl internal rotation molecule.}
\end{table*}

\section{Tentative detection of cis-HCOOH toward G+0.693-0.027}
\label{AnalysisHCOOH}

\begin{center}
\begin{figure*}[ht]
     \centerline{\resizebox{1.05
     \hsize}{!}{\includegraphics[angle=0]{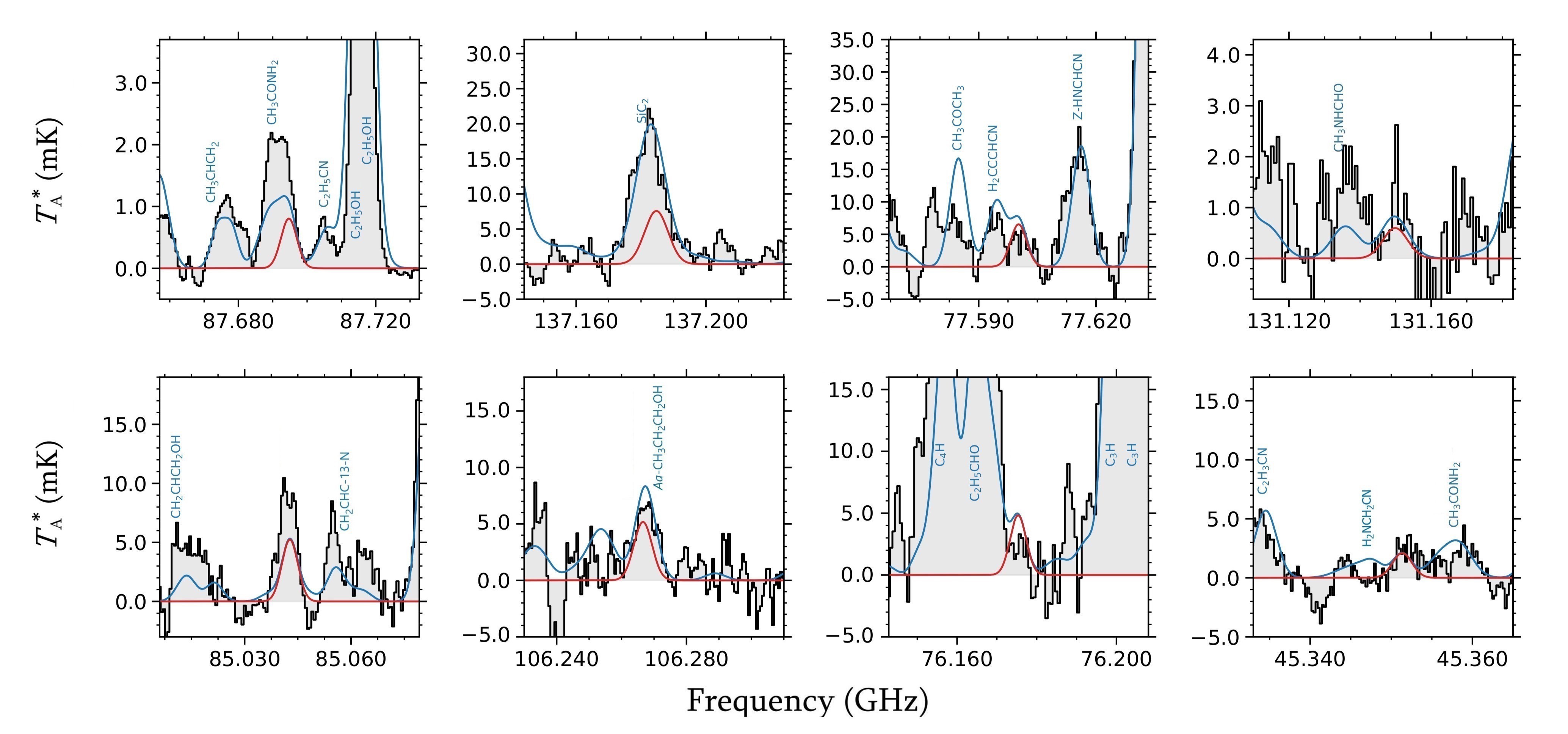}}}
     \caption{Selected transitions of $c$-HCOOH identified toward the G+0.693–0.027 molecular cloud. The best LTE fit computed with \textsc{Madcuba} is depicted in a red solid line, while the expected molecular emission from all of the molecular species identified in our spectral survey is shown in blue (observed spectra plotted as gray histograms).}
\label{f:LTEHCOOH}
\end{figure*}
\end{center}

We employed the rotational data of the ground vibrational state of $c$-HCOOH reported in \citet{Winnewisser2002}, which corresponds to the 046507 entry of the Cologne Database for Molecular Spectroscopy (CDMS) catalog \citep{Muller2005}. We have detected three lines with negligible contamination from other species and five additional lines exhibiting slight blends although the overall detection of this conformer is somewhat tentative. Nevertheless, many other lines that are predicted to be also very bright (e.g., the 5$_{0,5}$-4$_{0,4}$ and 2$_{0,2}$-1$_{0,1}$ transitions) could not be included in the current analysis since they are significantly blended with other molecular species. We thus used \textsc{Madcuba-Slim}, leaving in this case as free parameters the molecular column density ($N$) and the linewidth (FWHM), while the value of the excitation temperature ($T_{\rm ex}$) and the radial velocity ($v$$_{\rm LSR}$) were fixed in the fit to the values of $t$-HCOOH \citep{rodriguez-almeida2021a}. The derived physical parameters of $c$-HCOOH are listed in Table \ref{tab:comparisonacids}, and the results of the LTE fits of the lines are shown in Figure \ref{f:LTEHCOOH}. 

\begin{table*}
\centering
\tabcolsep 3pt
\caption{Spectroscopic information of the selected unblended or slightly blended transitions of \textit{c}-HCOOH detected toward G+0.693$-$0.027 (shown in Fig. \ref{f:LTEHCOOH}).}
\begin{tabular}{ccccccccc}
\hline\hline
Frequency & Transition $^{(a)}$ & log \textit{I}& \textit{g}$_u$ & E$_{LO}$ & E$_{up}$ & Comments  \\ 
(GHz) & &  (nm$^2$ MHz) & & (cm$^{-1}$) &  (K ) & & \\
\hline
45.3513500  & 2$_{1,1}-$1$_{1,0}$ & -4.7855 & 5 & 3.3 & 6.8 & Unblended  \\ 
76.1753218  & 1$_{1,0}-$1$_{0,1}$ & -4.3110 & 3 & 0.7 & 4.7 &  Blended with CH$_3$NHCHO \\ 
77.6001999  & 2$_{1,1}-$2$_{0,2}$ & -4.0803 & 5 &  2.2 & 6.8 &  Blended with H$_2$CCCHCN  \\
85.0427471  & 4$_{1,4}-$3$_{1,3}$ & -3.8501 & 9 &  6.8 & 13.7 &  Unblended  \\
87.6946938  & 4$_{0,4}-$3$_{0,3}$ & -3.7907 & 9 &  4.4 & 10.5 &  Blended with CH$_3$CONH$_2$ and U-line  \\
106.2665926 & 5$_{1,5}-$4$_{1,4}$ & -3.5561 & 11 &  9.6 & 18.8 &  Blended with $Aa$-CH$_3$CH$_2$CH$_2$OH  \\
131.1498866 & 6$_{0,6}-$5$_{0,5}$ & -3.2804 & 13 &  11.0 & 21.9 & Unblended   \\
137.1847188 & 3$_{1,3}-$2$_{0,2}$ & -3.6782 & 7  &   2.2  & 9.7 & Blended with SiC$_2$  \\
\hline 
\end{tabular}
\label{tab:LTEHCOOH}
\vspace*{-2.5ex}
\tablecomments{$^{(a)}$ The rotational energy levels are labelled using the conventional notation for asymmetric tops: $J_{K_{a},K_{c}}$, where $J$ denotes the angular momentum quantum number, and the $K_{a}$ and $K_{c}$ labels are projections of $J$ along the $a$ and $c$ principal axes.}

\end{table*}


\section{Complementary Figures}
\label{Figratio}

\begin{center}
\begin{figure}[ht]
     \centerline{\resizebox{1.05
     \hsize}{!}{\includegraphics[angle=0]{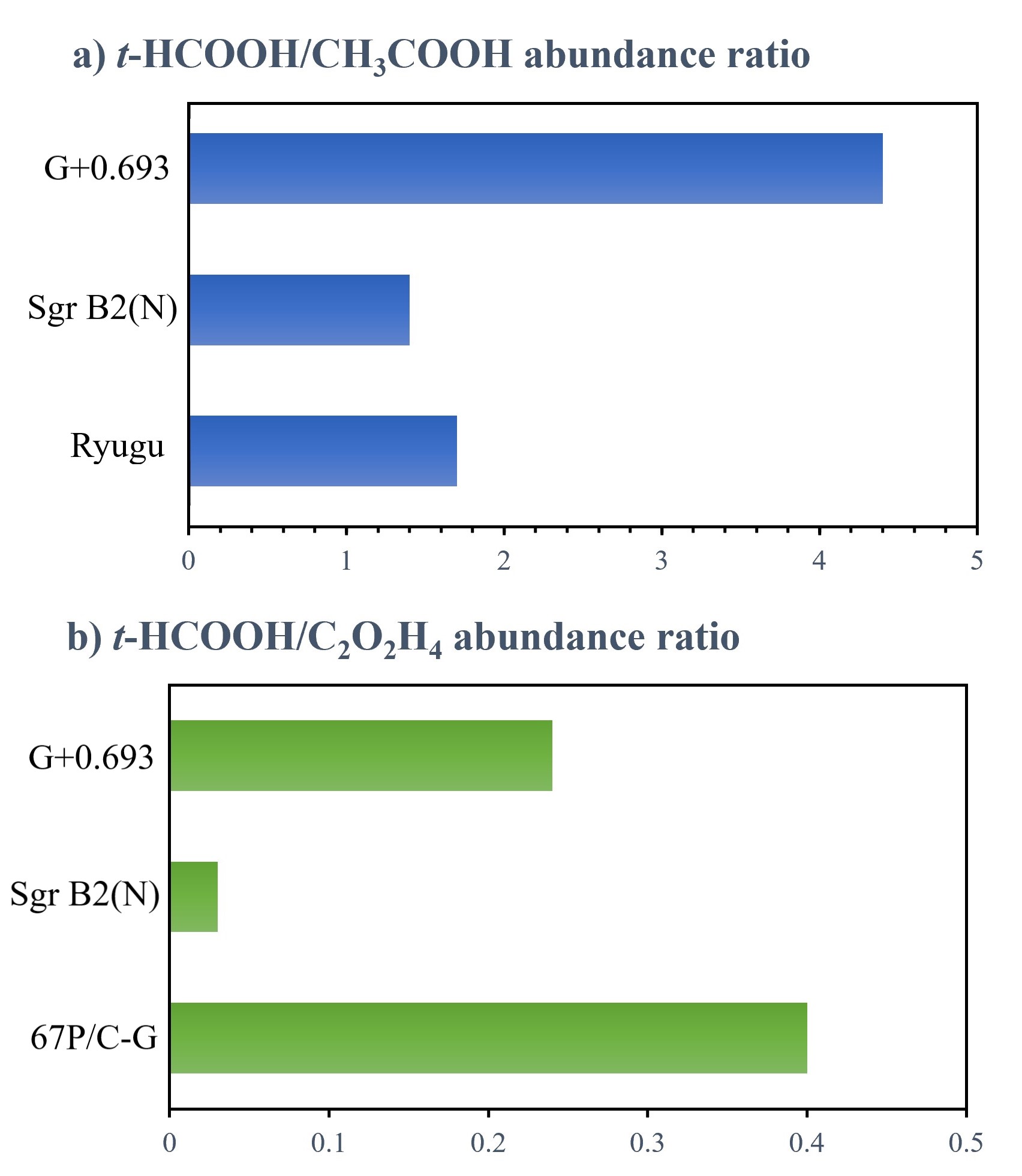}}}
     \caption{Abundance ratio between formic and acetic acid toward different astronomical environments}
\label{f:ratioref}
\end{figure}
\end{center}

In Figure \ref{f:ratioref} we provide a visual comparison of the abundance ratio between HCOOH and CH$_3$COOH toward different astronomical environments.

\end{document}